\documentclass[10pt,onecolumn]{article}
\usepackage[T1]{fontenc}

\usepackage{geometry}
\usepackage{graphicx}%
\usepackage{multirow}%
\usepackage{amsmath,amssymb,amsfonts,amsthm,mathrsfs}%
\usepackage{abstract}
\usepackage{xcolor}%
\usepackage{textcomp}%
\usepackage{manyfoot}%
\usepackage{booktabs, tabularx, nicematrix}%
\usepackage{algorithm}%
\usepackage{algorithmicx}%
\usepackage{algpseudocode}%
\usepackage{listings}%
\usepackage{siunitx}
\usepackage[hidelinks]{hyperref}   
\usepackage{cite}
\usepackage{authblk}

\usepackage{microtype}

\usepackage{titlesec}  
\titleformat{\section}[block]{\large\bfseries}{\thesection}{0.8ex}{}
\titlespacing{\section}{0pc}{2ex}{0.5ex}

\titleformat{\subsection}[block]{\bfseries}{\thesubsection}{0.8ex}{}
\titlespacing{\subsection}{0pc}{1ex}{0.5ex}

\titleformat{\subsubsection}[runin]{\bfseries}{\thesubsubsection}{0.8ex}{}
\titlespacing{\subsubsection}{0pc}{1ex}{0.5ex}

\usepackage[skip=3pt,labelsep=period]{caption}          % pie de foto o tabla

\makeatletter
\renewcommand{\maketitle}{\bgroup\setlength{\parindent}{0pt}
\begin{centering}
  \textbf{\Large{\@title}} \\ [4mm]
  \normalsize{\@author} \\[5mm]
  \normalsize{\@date} \\[1cm]
\end{centering}\egroup
}
\makeatother

\title{Boundary-dominated optomechanics in silicon metamaterial membranes}
\author[1,a,*,\dag]{David González-Andrade}
\author[1,b,*,\dag]{Paula Nuño Ruano}
\author[2]{Jianhao Zhang}
\author[1]{Paul Joseph Robin}
\author[1]{Hiba El Batoul Ferhat}
\author[1]{Samson Edmond}
\author[2]{Pavel Cheben}
\author[1]{Daniele Melati}
\author[1]{Eric Cassan}
\author[1]{Laurent Vivien}
\author[1]{Delphine Marris-Morini}
\author[1]{Norberto Daniel Lanzillotti-Kimura}
\author[1]{Carlos Alonso-Ramos}
\affil[1]{Centre de Nanosciences et de Nanotechnologies, CNRS, Université Paris-Saclay, 91120 Palaiseau, France}
\affil[2]{National Research Council Canada, 1200 Montreal Road, Bldg. M50, Ottawa, Ontario K1A 0R6, Canada}
\affil[a]{Current address: Telecommunication Research Institute (TELMA), Universidad de Málaga, E.T.S.I. Telecomunicación, 29010 Málaga, Spain}
\affil[b]{Current address: Photonic Systems Laboratory, École Polytechnique Fédérale de Lausanne, CH-1015 Lausanne, Switzerland}
\affil[*]{These authors contributed equally}
\affil[$\dag$]{Corresponding authors: \url{dgandrade@uma.es}, \url{paula.nuno-ruano@c2n.upsaclay.fr}}

\date{{\small(Dated: \today)}}

\newcommand{\Wm}{\unit{\per\W\per\m}}

\begin{document}
%TC:ignore
\twocolumn[
    \maketitle
    \renewcommand{\abstractname}{}
    \begin{onecolabstract}
        \noindent Stimulated Brillouin scattering in integrated photonic waveguides enables coherent coupling between optical photons and gigahertz acoustic phonons, providing a powerful mechanism for on-chip microwave photonics and opto-acoustic signal processing. Despite theoretical predictions of ultra-strong Brillouin interactions arising from enhanced light–sound coupling at device boundaries, most state-of-the-art integrated demonstrations remain governed by bulk photoelastic effects. This limitation stems from trade-offs between optical loss, interaction with waveguide boundaries and accessible phonon frequencies associated with the use of transverse-electric optical modes coupled to horizontally breathing mechanical modes. Here we demonstrate a new approach based on transverse-magnetic optical modes coupled to vertically breathing mechanical modes in suspended silicon membranes engineered with subwavelength metamaterial claddings. In this geometry, the interaction is dominated by the moving-boundary effect occurring at smooth top and bottom interfaces, while the phonon frequency is set primarily by the membrane thickness rather than its width. We observe forward Brillouin interactions at a record frequency of 12 GHz with a gain of 7200 \Wm\ and a mechanical quality factor of 620, yielding the highest Brillouin gain-to-quality-factor ratio reported in silicon waveguides. The devices exhibit net Brillouin amplification in millimeter-scale waveguides with pump powers below 15 mW, establishing a scalable platform for high-frequency integrated opto-acoustic signal processing.
        \medskip
        
        \noindent\textbf{Keywords:}  Integrated optomechanics, Forward stimulated Brillouin scattering, Moving boundary, Silicon photonics,  Subwavelength metamaterial membrane

        \bigskip 
    \end{onecolabstract}
]
%TC:endignore

%\section*{Introduction}
Stimulated Brillouin scattering (SBS) is a fundamental interaction between light and sound that enables coherent energy transfer between optical photons and acoustic phonons. By linking terahertz optical frequencies with gigahertz mechanical excitations, SBS provides a powerful mechanism for manipulating optical signals through acoustic waves \cite{Merklein_100yearsBS_2022}. This interaction underpins a wide range of technologies, from ultra-narrow linewidth lasers and distributed fiber sensing to microwave photonic signal processing, slow-light schemes, and non-reciprocal devices \cite{Gaeta_TunableDelayBSslowlight_2005, Boyd_storedLight_OpticalFibre_BS_2007, Feng_NoiseFree_filterBS_2018, Loh_ultranarrowBSlaser_nKref_2019, Soto_BSdistritibutedsensor_2021, Stiller_nonreciprocalVortexIsolatorBS_2022}. Extending Brillouin interactions to integrated nanophotonic platforms could unlock compact opto-acoustic processors and new forms of photon–phonon signal transduction \cite{eggleton_BrillouinIntegrated_2019}. However, achieving strong and controllable Brillouin coupling in nanoscale waveguides remains challenging because the mechanisms governing light–sound interactions differ fundamentally from those in conventional optical fibers.

In conventional fiber systems, Brillouin scattering is dominated by bulk nonlinear effects such as electrostriction and the photoelastic response of the medium. At the nanoscale, however, strong optical confinement dramatically enhances light-matter interaction at material interfaces, and radiation pressure at moving boundaries can significantly enhance optomechanical coupling. Theory predicts that nanoscale suspended silicon waveguides could support Brillouin interactions orders of magnitude stronger than those in silica fibers, opening the possibility of a boundary-dominated regime of Brillouin optomechanics \cite{Rakich_GiantBS_SWG_2012}. Despite these predictions, most demonstrations of Brillouin scattering in integrated photonic waveguides --including silicon \cite{kittlaus_large_2016, Otterstrom_BSsiLaser_2018}, silicon nitride \cite{Marpaung_BS_SiN_2022}, chalcogenide \cite{Pant_chalcogenideBBS_2011}, and thin-film lithium niobate \cite{Marpaung_IntegratedBS_tfln_2025} platforms-- remain dominated by the bulk photoelastic effect. This limitation arises largely from the modal configurations commonly employed, which rely on transverse-electric optical modes coupled to horizontally breathing mechanical modes. In this approach, strong optical interaction occurs at etched sidewalls, which introduce optical loss, imposing trade-offs between propagation loss, mechanical confinement, and accessible acoustic frequencies. Early silicon nanowires demonstrated efficient coupling and high gain-to-quality factor ratios in non-slow-light regimes \cite{van_laer_interaction_2015, van_laer_net_2015}. However, sidewall-mediated interactions in transverse-electric polarization heighten roughness sensitivity, and the small supporting pedestal limits scalability and platform robustness.

Here, we show a new regime of integrated Brillouin optomechanics in suspended silicon membranes, dominated by radiation pressure at moving boundaries. Our approach combines transverse-magnetic optical modes with vertically breathing mechanical modes and employs subwavelength metamaterial claddings to engineer optical and acoustic confinement independently. This geometry redirects the optomechanical interaction away from rough etched sidewalls toward the smooth top and bottom interfaces of the membrane, enabling strong boundary-mediated coupling while maintaining low optical loss. Using this architecture, we demonstrate forward Brillouin interactions at telecoms wavelengths with acoustic phonons at 12 GHz, the highest frequency reported in silicon Brillouin waveguides. The devices exhibit a Brillouin gain of 7200 \Wm\ and a mechanical quality factor of 620, yielding the largest reported ratio of gain-to-quality-factor in silicon systems. We observe net Brillouin amplification in millimeter-scale waveguides at pump powers below 15 mW and record an anti-Stokes depletion of $-26.8$ dB. The remarkably large gain reveals nonlinear dynamics not captured by conventional small-signal Brillouin models, in which optomechanical coupling generates additional anti-Stokes photons. We develop a theoretical description that predicts the possibility of simultaneous amplification of both Stokes and anti-Stokes fields and present preliminary experimental evidence of this regime in integrated waveguides.

\section*{Results}

\subsection*{Design}
\begin{figure*}[htbp]
    \centering
    \includegraphics[width=\textwidth]{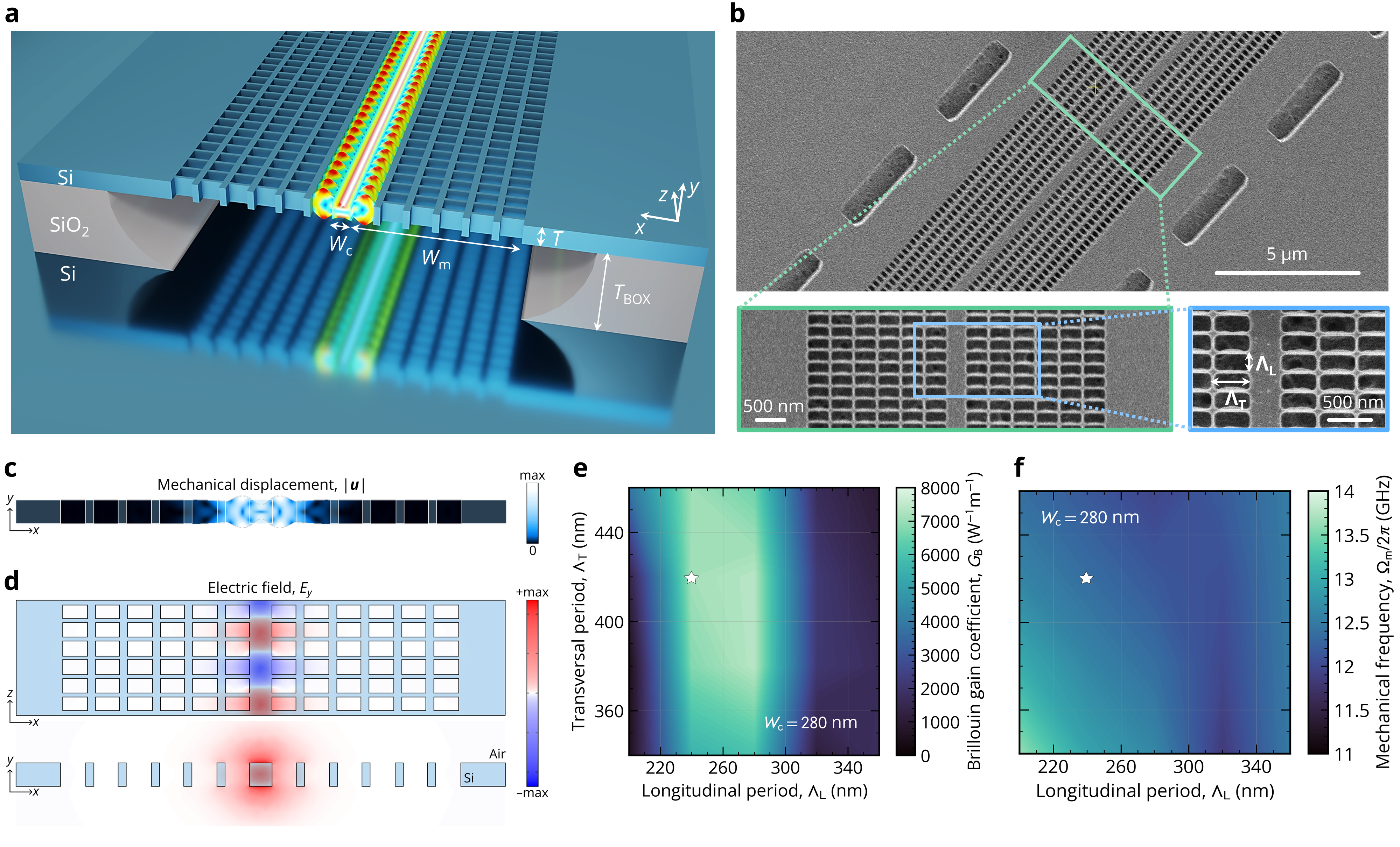}
    \caption{\textbf{Suspended Brillouin waveguide with lateral metamaterial membranes on a silicon-on-insulator platform.} (a) Schematic of the suspended optomechanical waveguide, showing the silicon core of width $W_\mathrm{c}$, the lateral metamaterial membranes of width $W_\mathrm{m}$, the silicon device layer thickness $T$, and the buried oxide (BOX) thickness $T_\mathrm{BOX}$. (b) Scanning electron microscope images of the fabricated membrane waveguide, with magnified views highlighting the longitudinal ($\Lambda_\mathrm{L}$) and transverse ($\Lambda_\mathrm{T}$) periods of the subwavelength membrane. Scale bars: 5 µm, 500 nm and 500 nm. (c) Simulated mechanical displacement magnitude of the guided acoustic mode (cross-section view). (d) Electric field distribution $E_\mathrm{y}$ of the optical mode, show in top view (top panel) and cross-section (bottom panel). Simulated (e) Brillouin gain coefficient $G_\mathrm{B}$ and (f) mechanical frequency $\Omega_\mathrm{m}/2\pi$ for a core width of $W_\mathrm{c} = 280$ nm as a function of the membrane periods $\Lambda_\mathrm{L}$ and $\Lambda_\mathrm{T}$. The star marks the fabricated geometry with $\Lambda_\mathrm{L} = 240$ nm and $\Lambda_\mathrm{T} = 420$ nm.}
    \label{fig:schematic}
\end{figure*}

Figure \ref{fig:schematic}a shows the subwavelength acoustic (SWA) membrane waveguide. The structure consists of a central waveguide core of width $W_\mathrm{c}$, laterally surrounded by a subwavelength metamaterial cladding of width $W_\mathrm{m}$ \cite{cheben_subwavelength_2018, Cheben_MetamaterialsIntegratedOptics_2023}. The cladding is defined by independent longitudinal ($\Lambda_\mathrm{L}$) and transverse ($\Lambda_\mathrm{T}$) periodicities (see Fig. \ref{fig:schematic}b, bottom right panel). This dual periodicity provides additional degrees of freedom for tailoring both optical and mechanical confinement, enabling precise control of evanescent fields and enhanced spatial overlap between optical and acoustic modes. At the same time, the metamaterial cladding ensures mechanical robustness, allowing the membrane to be suspended over centimeter-scale lengths \cite{Dihn_ControlModeConfine_2023}. The crystalline silicon device layer and buried oxide thicknesses are 300 nm and 3 µm, respectively, consistently with standard silicon-on-insulator wafers available in our cleanroom. Scanning electron microscope (SEM) images of the fabricated devices are shown in Fig. \ref{fig:schematic}b, together with magnified views highlighting the dual-periodic membrane structure.

We focus on intramodal forward stimulated Brillouin scattering between the guided fundamental quasi-transverse-magnetic (TM) optical mode and an out-of-plane breathing mechanical mode. Optical confinement is provided by the refractive-index contrast between the waveguide core and the surrounding metamaterial cladding, while acoustic confinement arises from the large impedance contrast between silicon and air. A key advantage of this vertical optomechanical configuration is the strong suppression of phonon leakage through lateral supports, as the dominant mechanical motion is perpendicular to the chip plane (see Fig. \ref{fig:schematic}c). This geometry also mitigates optical propagation losses. For the quasi-TM mode, the dominant electric-field component is oriented normal to the waveguide plane, resulting in reduced interaction with etched sidewalls as shown in Fig. \ref{fig:schematic}d. Because reactive-ion etching (RIE) introduces stochastic roughness primarily at the sidewalls, the SWA waveguide is intrinsically less susceptible to scattering losses than conventional Brillouin designs based on transverse-electric (TE) polarization. By contrast, top and bottom interfaces (accessed only during the release step) remain comparatively smooth, further contributing to low-loss optical guidance.

To optimize the SWA membrane geometry, we first constrained the minimum dimensions of the silicon tethers within the subwavelength membrane to 50 nm along the longitudinal direction and 100 nm along the transverse direction. These dimensions ensure mechanical robustness while minimizing residual lateral phonon leakage by reducing the amount of silicon available for acoustic radiation \cite{GonzalezAndrade_SimuTMforBS_2024}. With these constraints in place, the membrane periodicities were judiciously adjusted to maximize the Brillouin gain coefficient $G_\mathrm{B}$ and reduce optical leakage towards the silicon substrate (Supplementary Information, sections A and B). The first optimization exploits the additional degrees of freedom provided by the dual-periodic cladding, enabling fine control over the photon-phonon overlap through simultaneous tuning of the waveguide core and the longitudinal ($\Lambda_\mathrm{L}$) and transverse ($\Lambda_\mathrm{T}$) periods, while remaining within the subwavelength regime to suppress Bragg and diffraction effects (Supplementary Information, section C). Representative simulation results for a core width of $W_\mathrm{c}=280$ nm are shown in Figs. \ref{fig:schematic}e and \ref{fig:schematic}f. The structure is single-mode for the optical field while supporting multiple mechanical modes. Across the explored parameter space, the vertically breathing acoustic mode exhibits frequencies between 11.85 and 13.59 GHz. Notably, Brillouin gain coefficients exceeding 4000 \allowbreak \Wm\ are obtained over a broad range of longitudinal periods $\Lambda_\mathrm{L}$ (210 -- 300 nm), largely independent of the transverse period $\Lambda_\mathrm{T}$. This behavior indicates that variations in the transverse hole width have only a minor influence on vertical optical confinement, whereas tuning the longitudinal periodicity provides an effective handle for engineering mode confinement \cite{Dihn_ControlModeConfine_2023}.

\begin{figure*}[htb]
    \centering
    \includegraphics[width=\textwidth]{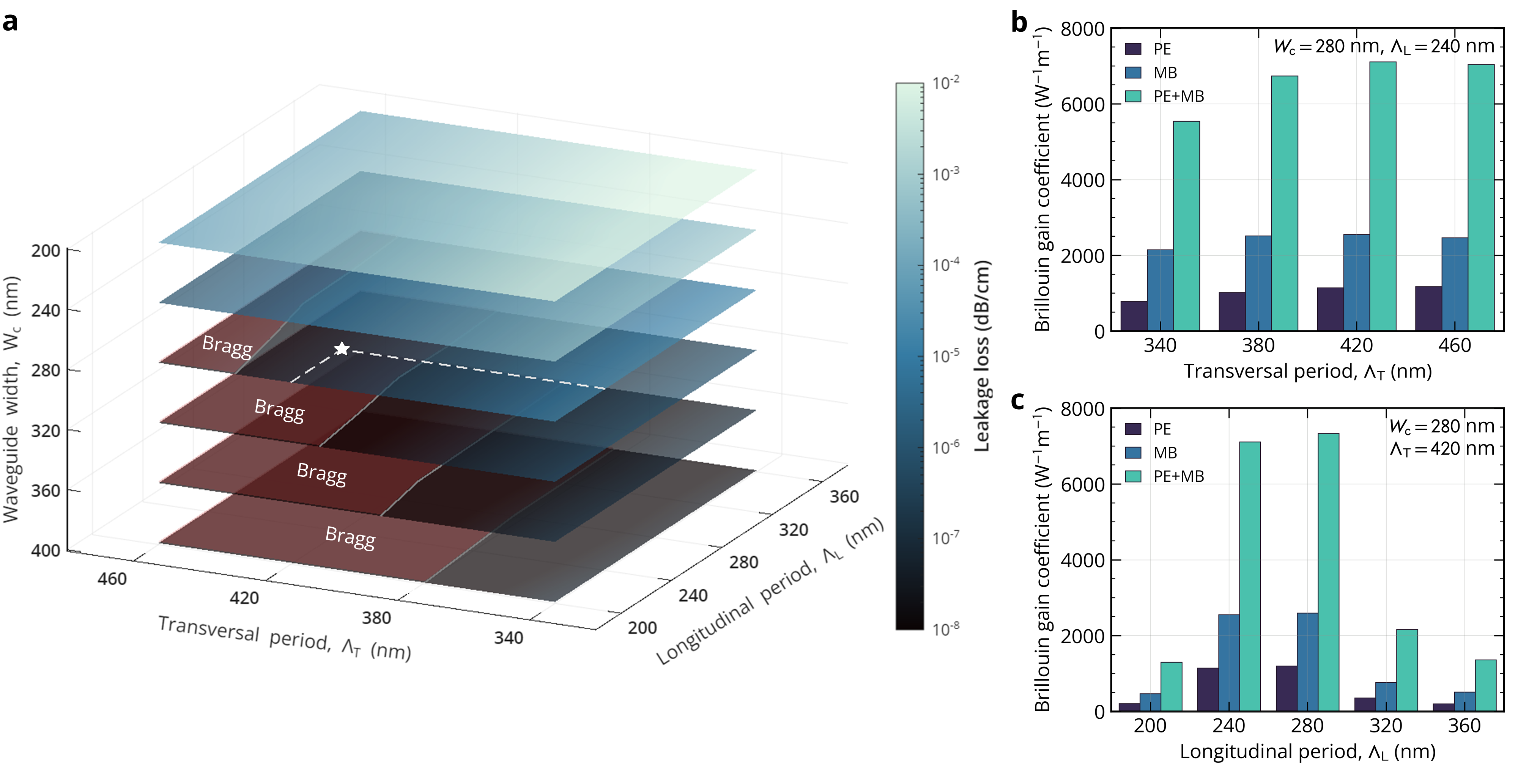}
    \caption{\textbf{Design landscape and simulated Brillouin performance.} (a) Three-dimensional map of the simulated optical leakage loss (dB/cm) as a function of the transverse ($\Lambda_\mathrm{T}$) and longitudinal ($\Lambda_\mathrm{L}$) membrane periods for several cored widths $W_\mathrm{c}$. For each surface, the Bragg region for the optical mode associated with the transverse membrane period is highlighted in red. The longitudinal Bragg condition lies far outside the explored parameter range, as the membrane operates in the subwavelength regime, and is therefore not shown. The star marks the parameters of the fabricated device: $W_\mathrm{c} = 280$ nm, $\Lambda_\mathrm{L} = 240$ nm, and $\Lambda_\mathrm{T} = 420$ nm. Simulated Brillouin gain coefficient of the selected geometry as a function of (b) $\Lambda_\mathrm{T}$ and (c) $\Lambda_\mathrm{L}$, along with the decomposed photoelastic (PE) and moving-boundary (MB) contributions. The total optomechanical interaction exceeds the sum of the individual PE and MB contributions due to coherent interference between the two mechanisms.}
    \label{fig:leakage}
\end{figure*}

In addition to maximizing photon-phonon coupling, optical leakage losses must be carefully managed, particularly because strong overlap favors narrow core widths below 400 nm \cite{Halir_lossesSWG_2016}. Figure \ref{fig:leakage}a maps the leakage loss as a function of core width and membrane periodicities. The red shaded region highlights transverse periods that approach the Bragg regime, where additional reflection losses emerge and should be avoided. Narrow cores (200 –- 240 nm) suffer from appreciable leakage, while wider cores (300 –- 400 nm) increase confinement of the optical mode and limit the achievable overlap (Supplementary Information, section B). An optimal compromise is found for $W_\mathrm{c}=280$ nm, for which the leakage loss remains below $2\cdot
10^{-5}$ dB/cm across all periodicity combinations. The relative contributions of the photoelastic (PE) and moving boundary (MB) effects to the Brillouin gain are shown in Figs. \ref{fig:leakage}b and \ref{fig:leakage}c for $W_\mathrm{c}=280$ nm. In all cases, the MB contribution dominates, underscoring that the SWA membrane design primarily enhances Brillouin coupling through strong boundary motion and efficient optical field localization at material interfaces, rather than through volumetric strain alone.

The waveguide geometry selected for fabrication is indicated by the star symbols in Figs. \ref{fig:schematic} and \ref{fig:leakage}, corresponding to $W_\mathrm{c}=280$ nm, $\Lambda_\mathrm{T}=420$ nm and $\Lambda_\mathrm{L}=240$ nm. This configuration supports a vertically breathing mechanical mode at 12.368 GHz and yields a simulated Brillouin gain coefficient of $G_\mathrm{B}=7110$ \Wm. Importantly, the SWA membrane is compatible with a streamlined fabrication flow based on a single, uniform etch step, avoiding the need for loading-effect engineering previously used to tailor acoustic confinement \cite{lei_antiresonant_2024}. This simplified fabrication flow improves reproducibility across millimeter- or centimeter-scale devices. The robustness of the design against fabrication-induced variation was systematically assessed, and a detailed tolerance analysis is provided in Supplementary Information, section D.

\subsection*{Experimental characterization}
To experimentally validate this design, we fabricated SWA membranes with lengths of 4 and 6 mm using standard silicon photonics procedures (see Methods). The devices exhibit linear losses ranging from 1.9 to 2.6 dB/cm. We characterized the stimulated Brillouin scattering in these waveguides using two- and three-tone measurements schemes (see Supplementary Information, sections E and F).

\subsubsection*{Three-tone pump configuration.} 
\begin{figure}[htb]
    \centering
    \includegraphics[width=\columnwidth]{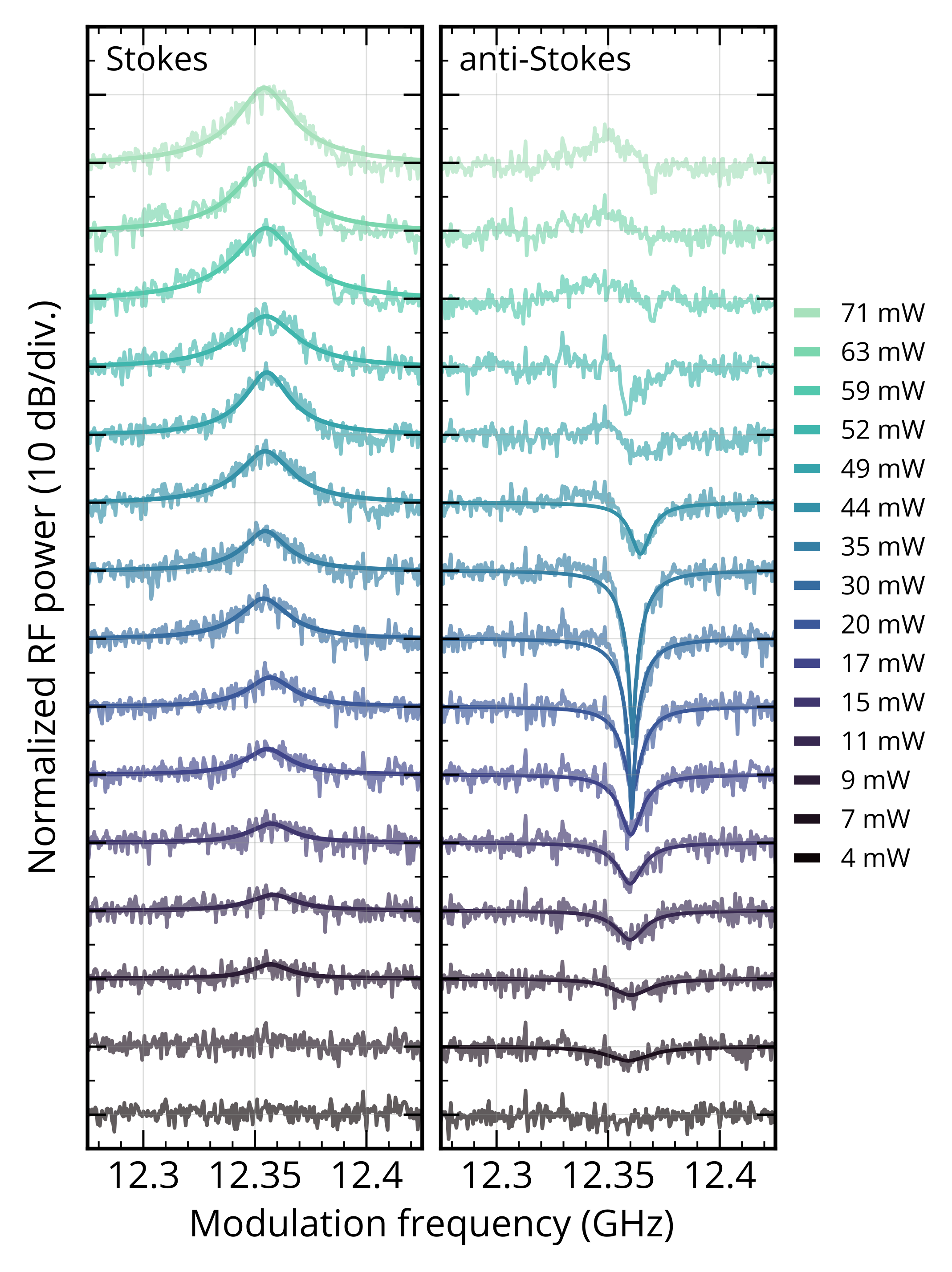}
    \caption{\textbf{Optomechanical spectra in a three-tone pump experiment.} RF beat note between the reference and the Stokes (left panel) and anti-Stokes (right panel) lines as a function of the modulating frequency for different input pump powers for a 6 mm-long waveguide.}
    \label{fig:3tone_RF_power}
\end{figure}

We modulate the narrow-linewidth pump light to generate Stokes and anti-Stokes sidebands. The three lines are then coupled into the waveguide via integrated grating couplers. In Fig. \ref{fig:3tone_RF_power}, we show the normalized sideband power as a function of the modulating frequency for different on-chip pump powers. We observe a clear signature of the optomechanical effect when the modulating frequency sweeps across the acoustic resonance. After fitting the data to the theoretical model (Supplementary Information, section G), we obtain a mechanical frequency of $\Omega_\mathrm{m}/2\pi = 12.358 \pm 0.003$ GHz, in excellent agreement with the expected value ($\Omega_\mathrm{m}/2\pi = 12.368$ GHz), and a linewidth of $\gamma_\mathrm{m} = \Gamma_\mathrm{m}/2\pi = 20 \pm 2$ MHz, corresponding to a mechanical quality factor of $Q_\mathrm{m} \sim 620$ and a phonon lifetime of $\tau_\mathrm{m} = 1/\gamma_\mathrm{m} \sim 8$ ns. This acoustic frequency is among the largest reported to date for forward Brillouin scattering in silicon membranes (see Table \ref{tab:comparison_state_art}).

\begin{figure}[htb]
    \centering
    \includegraphics[width=\columnwidth]{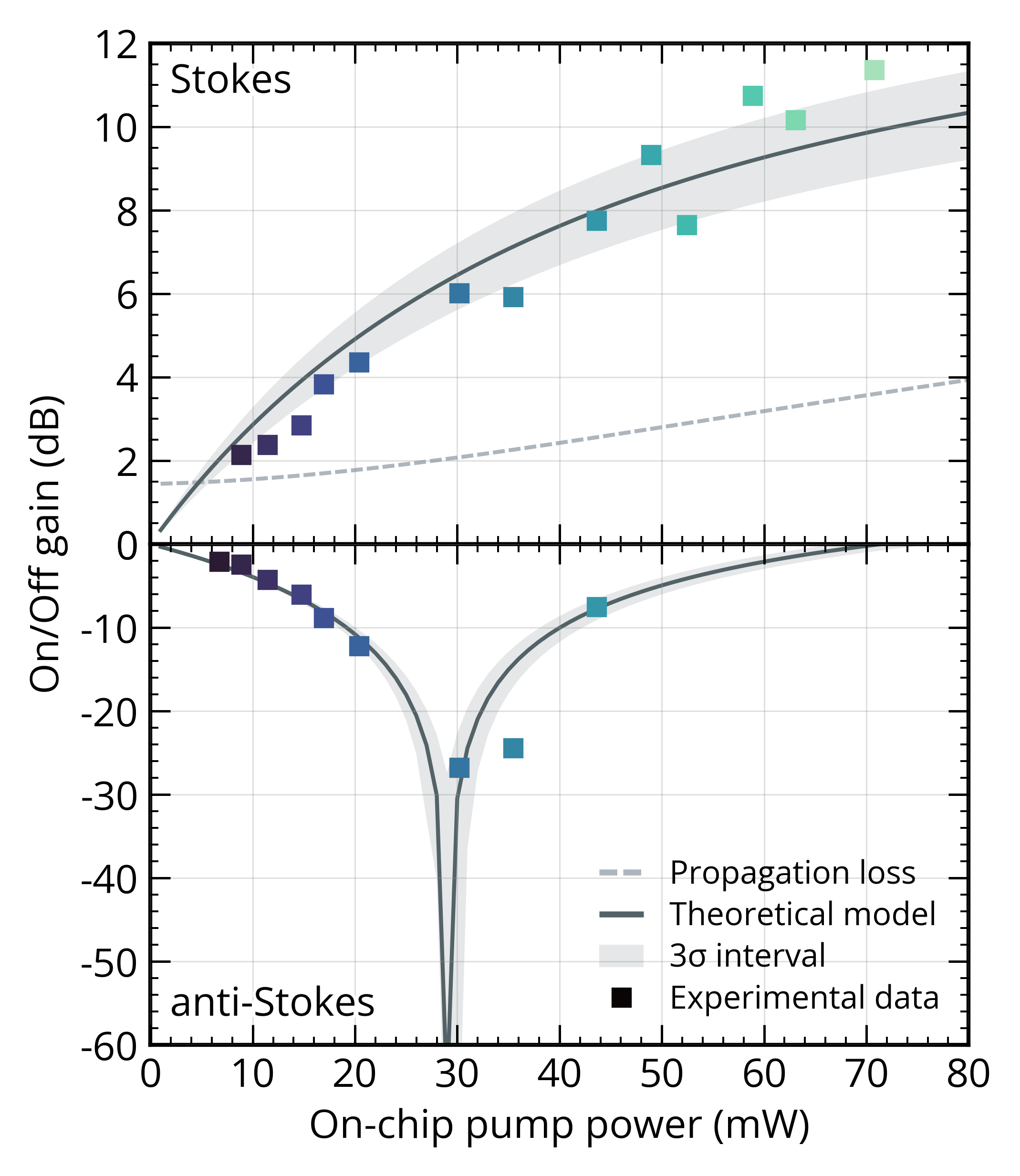}
    \caption{\textbf{Optomechanical response in the three-tone pump experiment.} On/off gain for the Stokes (upper panel) and anti-Stokes (lower panel) lines at the output of the waveguide as a function of the input pump power for a 6 mm-long waveguide. The squares correspond to the experimental data, the solid line refers to the theoretical model, and the dashed line represents the total propagation loss including the linear and nonlinear contributions. The shaded region highlights the $3\sigma$ confidence interval.}
    \label{fig:3tone_onoff}
\end{figure}

Figure \ref{fig:3tone_onoff} depicts the on/off gain, i.e., the relative variation in the sideband power, as a function of the input power. By fitting the experimental data to the theoretical model, we can estimate the Brillouin gain coefficient $G_\mathrm{B}$ of the waveguide. From the Stokes measurements, we obtain a gain coefficient of $G_\mathrm{B} = 7717 \pm 453$ \Wm, and from the anti-Stokes data, a value of $G_\mathrm{B} = 7211 \pm 102$ \Wm, with an associated $G_\mathrm{B}/Q_\mathrm{m} \sim 12 \pm 0.4$. The difference in the values may be due to uncertainties in the estimates of the initial pump, Stokes, and anti-Stokes powers, arising from instabilities in the input fiber and thermal drift of the electro-optic modulator. These values agree well with the simulated one ($G_\mathrm{B} = 7110$ \Wm), with the difference coming from uncertainties in the material parameters and small fabrication deviations from the nominal design.

Our design yields the strongest $G_\mathrm{B}/Q_\mathrm{m}$ measured in silicon waveguides (see Table \ref{tab:comparison_state_art}) due to the strong moving-boundaries effect as shown in Figs. \ref{fig:leakage}b and \ref{fig:leakage}c. The $G_\mathrm{B}/Q_\mathrm{m}$ can be understood as the waveguide equivalent for the single photon-phonon coupling \cite{laer_BSoptomech_2016} and is often used as a metric to compare different geometries with variable mechanical confinement. These results are consistent across measured waveguides of different core widths and lengths, as shown in section H of the Supplementary Information. 

In Fig. \ref{fig:3tone_onoff} we depict the relative power variations of the sidebands' powers due to the Brillouin effect. The Stokes line (upper panel) shows a monotonic increase with the pump power. Using the three-tone pump scheme, net Stokes gain is experimentally achieved with pump powers as low as 8 mW, with a theoretical threshold at 5 mW and a maximum of 7.8 dB of net gain at 70 mW. The anti-Stokes mode (lower panel) shows two distinct trends. At low pump powers, the loss increases until the anti-Stokes line is fully depleted due to energy transfer to the pump. We measure up to $-26.8$ dB of depletion for an input power of 30 mW. To the best of our knowledge, this is the largest anti-Stokes depletion achieved to date in a silicon waveguide using purely the optomechanical effect. Comparable suppression values have been achieved before using complex techniques such as radio frequency (RF) cancellation \cite{bedoya_notch_filter_pedestral_2015} or multi-\allowbreak pole photonic–\allowbreak phononic emitter–\allowbreak receivers \cite{Gertler_RFfilterSi_2020, gertler_MWP_RFfilter_2wg_2022}. For pump powers above this value, the optomechanical effect depletes the input anti-Stokes line somewhere along the waveguide and starts generating new anti-Stokes photons. As a result, a lower on/off loss and a weaker spectral peak are observed experimentally (e.g., see right panel in Fig. \ref{fig:3tone_RF_power} for an input power of 44 mW). Interestingly, for sufficiently large pump powers, the model predicts a positive on/off gain for the anti-Stokes line, resulting in the amplification of the anti-Stokes mode (see section G of the Supplementary Information). The weak amplification peaks for input powers above 60 mW in the right panel of Fig. \ref{fig:3tone_RF_power} constitute a preliminary proof of this effect. We observed similar features in other waveguides we measured (see Supplementary Information, section H). However, additional measurements would be valuable for corroborating these results.  

This work provides the first observation of the simultaneous generation of Stokes and anti-Stokes photons in an integrated waveguide. However, due to the instability of the input fiber at high pump powers, a more detailed study of the simultaneous amplification of Stokes and anti-Stokes modes was not possible.

\subsubsection*{Two-tone pump configuration.} 
\begin{figure}[htb]
    \centering
    \includegraphics[width=\columnwidth]{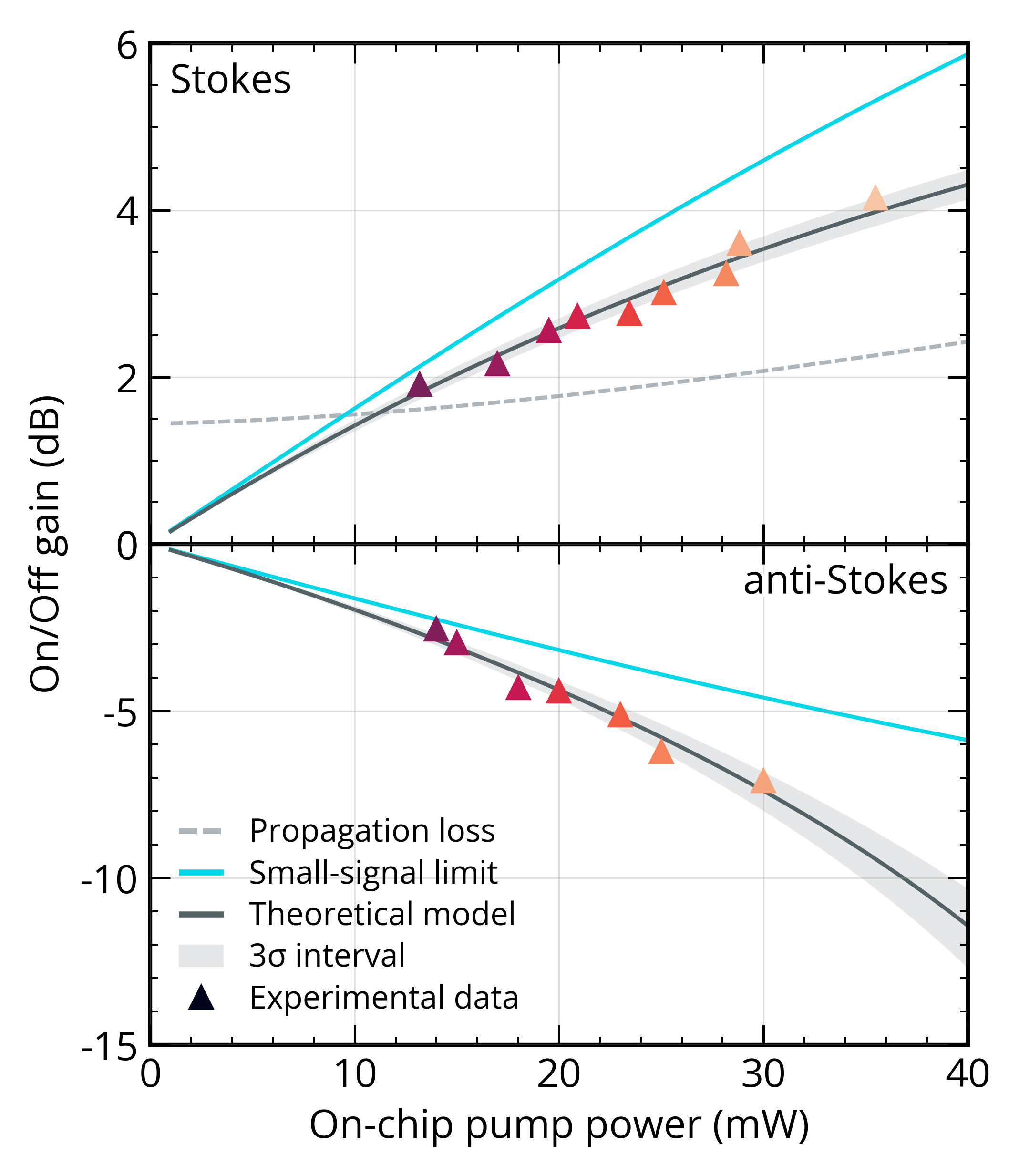}
    \caption{\textbf{Optomechanical response in the two-tone pump experiment.} On/off gain for the Stokes (upper panel) and anti-Stokes (lower panel) modes at the output of the waveguide as a function of the input pump power for a 6 mm-long waveguide. The triangles correspond to the experimental data, the solid gray line refers to the theoretical model, the solid light blue line refers to the small-signal limit, and the dashed line represents the total propagation loss including the linear and nonlinear contributions. The shaded region highlights the $3\sigma$ confidence interval.}
    \label{fig:2tone_onoff}
\end{figure}

In Fig. \ref{fig:2tone_onoff}, we depict the relative power variation for the Stokes and anti-Stokes lines (top and bottom panels, respectively) measured using the standard two-tone pumping scheme, in which one of the sidebands is filtered out before injecting the light into the waveguide. The experiment shows a clear asymmetry between the Stokes and anti-Stokes variation due to the non-exponential energy transfer in intramodal forward Brillouin scattering, previously reported only when using the three-tone pump configuration \cite{BS_3toneGain_2021, PNR_SWGPhnC_2024}. Traditional experiments in Brillouin scattering in integrated waveguides focus only on the Stokes evolution and overlook the asymmetry between the Stokes gain and the anti-Stokes depletion. To date, all two-tone pump experiments have been analyzed assuming exponential energy transfer between the pump and the sideband. This model, often referred to as the \textit{small-signal limit} (light blue curve in Fig. \ref{fig:2tone_onoff}), predicts the same response, albeit the sign, for the Stokes and the anti-Stokes power. However, it neglects the coupling between the sidebands characteristic of intramodal FBS (see Supplementary Information, section G) \cite{Wolff_Chapter2_theorySBS_2022} and fails to explain the experimental asymmetry between Stokes gain and anti-Stokes depletion.

Using the two-tone pump scheme, we measure a lower on/off response than in to the three-tone pumping configuration, due to the gain enhancement in the latter. A maximum on/off gain of 3.6 dB for the Stokes and 7.1 dB of loss for the anti-Stokes are achieved with 30 mW of input power. Net gain is obtained for input powers above 13 mW, with a maximum of 1.9 dB at 35 mW of input power, the maximum we used in this experiment. By fitting the experimental values to the theoretical model, we estimate a Brillouin gain coefficient of $G_\mathrm{B}=7019 \pm 123$ \Wm\ using the Stokes data and $G_\mathrm{B}=7764 \pm 129$ \Wm\ with the anti-Stokes data, comparable to the three-tone pumping estimations. As in the previous experiment, the difference in the values may be due to uncertainties in the estimation of the initial pump, Stokes, and anti-Stokes powers arising from fiber instabilities and thermal drift of the electro-optic modulator and the optical filter. Full depletion and regeneration of anti-Stokes photons are also to be expected in the two-tone pump scenario. However, longer waveguides and higher input powers than those used in this work would be needed to observe this behavior.

\section*{Discussion}
Our work demonstrates a new regime of integrated Brillouin optomechanics in suspended silicon membranes, where the interaction is dominated by the moving-boundary effect rather than the conventional bulk photoelastic response. By combining transverse-magnetic optical modes with vertically breathing mechanical modes and employing subwavelength metamaterial claddings, we redirect the optomechanical interaction to smooth top and bottom interfaces, reducing optical loss and enabling efficient photon–phonon coupling. This geometry also allows the phonon frequency to be determined primarily by the membrane thickness rather than the waveguide width, providing access to high-frequency phonons that were inaccessible in prior transverse-electric, horizontally breathing mode implementations.

A fundamental trade-off among optical loss, mechanical confinement, and accessible phonon frequencies has constrained the performance of previous integrated Brillouin platforms. In standard TE-horizontal geometries, strong light–sound interactions occur at etched sidewalls, which are inherently lossy, and the mechanical frequencies scale inversely with the waveguide width, limiting the Brillouin shift range. Our approach overcomes these constraints by establishing a boundary-dominated regime of SBS that enables both high coupling strength and high-frequency operation within compact, low-loss devices.

Experimentally, we observe Brillouin interactions at 12 GHz, the highest phonon frequency reported to date in silicon waveguides. The devices achieve a Brillouin gain of 7200 \Wm\ with a mechanical quality factor of 620, yielding the largest gain-to-linewidth ratio measured in integrated silicon systems. This high Brillouin shift in a forward configuration has only been reported previously in hybrid silicon/silicon nitride waveguides with much lower Brillouin gain coefficients \cite{Shin_SiSiN_2013} (see Table \ref{tab:comparison_state_art}). Our design yields the largest gain-frequency product to date for silicon optomechanical waveguides, $G_\mathrm{B} \cdot \Omega_\mathrm{m}/2\pi > 88920$ (\Wm) $\cdot$ \unit{\GHz}. We further demonstrate net Brillouin amplification in millimeter-scale waveguides with pump powers below 15 mW, without the need for extremely low losses. Previous results had only achieved a low pump power threshold for optical losses below 0.3 dB/cm, a limit hard to attain in integrated silicon waveguides, and lower mechanical frequencies due to the large waveguide width required for ultra-low loss optical propagation. Additionally, using a three-tone pumping scheme, we achieve a maximum attenuation of 27 dB for the anti-Stokes line with a pump power of 30 mW, highlighting the efficiency and scalability of the approach. Up to now, this level of attenuation based on the Brillouin scattering was only achieved using complex RF techniques \cite{bedoya_notch_filter_pedestral_2015, Gertler_RFfilterSi_2020, gertler_MWP_RFfilter_2wg_2022}.

Moreover, we experimentally observe a regime beyond the predictions of conventional small-signal models, in which the anti-Stokes sideband is fully depleted and new anti-Stokes photons are generated, leading to the simultaneous amplification of both Stokes and anti-Stokes lines at high pump powers. These observations highlight the necessity of using a full theoretical model that accounts for the intrinsic coupling between Stokes and anti-Stokes processes in intramodal forward Brillouin scattering, even when only a single sideband is initially excited \cite{Wolff_Chapter2_theorySBS_2022}. With improved coupling efficiency and more stable light injection, future experiments could explore the simultaneous achievement of net gain for both Stokes and anti-Stokes modes, as well as the generation of higher-order sidebands, opening new opportunities for complex opto-acoustic signal manipulation in integrated photonic systems.

Our results establish a versatile platform for high-frequency integrated opto-acoustic signal processing, with potential applications in on-chip microwave photonics, non-reciprocal devices, and coherent light storage. More broadly, the demonstration of a moving-boundary dominated SBS regime in silicon nanophotonics suggests that careful engineering of optical and mechanical modes can unlock previously inaccessible optomechanical phenomena. Future work could explore further optimization of membrane geometries, higher-frequency operation, and integration with other photonic functionalities, potentially enabling fully reconfigurable, compact opto-acoustic circuits.

\begin{table*}[hbtp]
    \centering
    \caption{Comparison with previous demonstrations of intramodal forward Brillouin scattering in suspended silicon waveguides. All data refer to a two-tone pump configuration. \\%
    \textsuperscript{*}This demonstration corresponds to a heterogeneous waveguide with a silicon core and a silicon nitride cladding. \\%
    \textsuperscript{**}This threshold corresponds to a simulation based on the experimental value of $G_\mathrm{B}$. The experimental data were collected using a three-tone pump scheme.
    }
    \label{tab:comparison_state_art}
    {\small
    \begin{NiceTabular}{@{}*{7}{c}@{}}
    \toprule
    \Block{2-1}{Ref.} & $\alpha$ & $\Omega_\mathrm{m}/2\pi$ & $G_\mathrm{B}/Q_\mathrm{m}$ & $L$ & \Block{2-1}{Stokes} & \Block{2-1}{anti-Stokes} \\
    & [dB/cm] & [GHz] & [\Wm] & [cm] & & \\
    \midrule
    \Block{}{Nat. Comm. 4 \\ (2013) \cite{Shin_SiSiN_2013}} & 7 & 1.8 -- 16.3\textsuperscript{*}  & 1.5 -- 2.8 & 0.33 & -- & \Block{}{-0.45 dB \\ (20 mW)} \\[5mm]
    \Block{}{New J. Phys. 17, 11 \\ (2015) \cite{van_laer_net_2015}} & 5 & 9.1 & 10.3 & 0.25 & \Block{}{Net gain: \\ $P_\mathrm{p}>30$ mW} & -- \\[5mm]
    \Block{}{Nat. Phot. 9, 3 \\ (2015)\cite{van_laer_interaction_2015}}& 2.6 & 9.2 & 10.5 & 4 & -- & \Block{}{-4.4 dB \\ (35 mW)} \\[5mm]
    \Block{}{Nat Phot. 10, 7 \\(2016) \cite{kittlaus_large_2016}} & 0.18 & 4.35 & 1.7 & 2.9 & \Block{}{Net gain: \\ $P_\mathrm{p}>5$ mW} & \Block{}{-7 dB \\ (36 mW)} \\[5mm]
    \Block{}{Nat. Comm. 15 \\ (2024) \cite{lei_antiresonant_2024}} & 0.3 & 3.6 -- 8.6& 2.5 -- 4.2 & 2.5 & \Block{}{Net gain: \\ $P_\mathrm{p}>5$ mW\textsuperscript{**} } & -- \\[5mm]
    \Block{}{This work \\ (two-tone pumping)} & 2 & 12.35 & 12 & 0.6 & \Block{}{Net gain: \\ $P_\mathrm{p}>13$ mW} & \Block{}{-7.1 dB \\ (30 mW)} \\
    \bottomrule
    \end{NiceTabular}
    }
\end{table*}

%TC:ignore
\section*{Methods}
\subsection*{Device fabrication.} 
SWA membrane waveguides were fabricated on SOI wafers with a 300-nm \{100\}-oriented silicon layer and a 3-µm buried oxide. All structures were defined using a single lithography and etching step. A positive-tone electron-beam resist (ZEP 520A) was deposited by spin coating and patterned using electron-beam lithography (Raith EBPG 5200, 100 keV acceleration voltage, 0.8 nA beam current, and a dose of 300 µC cm$^{-2}$). After exposure, the resist was developed in AR 600-546 for 40 s and cleaned in isopropyl alcohol (IPA), defining the waveguide core and dual-periodic membrane cladding. The pattern was transferred into the silicon device layer by inductively coupled plasma reactive ion etching (ICP-DRIE SPTS Rapier) using an SF$_6$/C$_4$F$_8$ chemistry at 0 $^\circ$C. Following etching, the samples were sequentially cleaned in butanone (CH$_3$CCH$_2$CH$_3$), a piranha solution (H$_2$SO$_4$:H$_2$O$_2$ = 3:1), and O$_2$ plasma (Stripper DIENER Nano) to remove residual resist and organic contamination. For suspended devices, the buried oxide beneath the patterned regions was removed using vapour-phase hydrofluoric acid (HF) etching for 1 h, releasing the silicon membrane.

\subsection*{Finite element simulations.} 
Optical and mechanical modes sustained by the structure are calculated using the finite-element commercial software COMSOL Multiphysics. The waveguide material is considered as optically isotropic, with refractive index $n_\mathrm{Si} = 3.45$, and elastically anisotropic, with elastic constants $[c_{11}, \allowbreak c_{12}, \allowbreak c_{44}] = [165.64,\allowbreak 63.94,\allowbreak 79.51]$ \unit{\GPa} and photoelastic constants $[p_{11}, p_{12}, p_{44}] = [-0.094,\allowbreak 0.017,\allowbreak -0.051]$. The waveguides are oriented along the $\mathbf{\hat{z}}$-axis, with the $\mathbf{\hat{y}}$-axis perpendicular to the chip plane. The fabricated waveguides are aligned along the crystallographic direction $\langle 110 \rangle$, so a rotation of $\theta=\pi/4$ around the $\mathbf{\hat{y}}$-axis is applied to the elastic and photoelastic tensors of silicon. A more detailed explanation can be found in the Supplementary Information, section A.

\subsection*{Experimental characterization.} 
SWA membrane waveguides are characterized using heterodyne measurements. The Stokes and anti-Stokes sidebands are obtained by modulating the pump line and sent into the waveguide in the three-tone pump scheme. In the two-tone pump scheme, one of the sidebands is filtered out before entering the waveguide. At the output, a reference beam is added by up-shifting the pump, and the beating note between the lines is recorded by a fast photodiode. More details, as well as the schematics of the experimental setups, can be found in the Supplementary Information, section F. 

\section*{Supplementary information}
Supplementary Information is available.

\section*{Acknowledgments}
In memory of Xavier Le Roux, for his invaluable work in developing and optimizing the fabrication processes at our cleanroom that laid the foundations for this work.

\section*{Declarations}
\subsection*{Funding} 
This work has been funded by the European Union’s Horizon Europe (Marie SklodowskaCurie grant agreement N° 101062518) and the European Union (ERC SPRING, grant agreement N° 101087901). This publication is part of the grant RYC2023-045670-I funded by MICIU/AEI/10.13039/501100011033 and by ESF+. This work was done within the C2N micro nanotechnologies platforms and partly supported by the RENATECH network and the General Council of Essonne. Views and opinions expressed are however those of the authors only and do not necessarily reflect those of the European Union or the European Research Executive Agency (REA). Neither the European Union nor the granting authority can be held responsible for them.

\subsection*{Competing interests} 
The authors declare no competing interests.

\subsection*{Ethics approval and consent to participate} 
Not applicable.

\subsection*{Consent for publication} 
Not applicable.

\subsection*{Data availability} 
The data supporting this study's findings are available from the corresponding author upon reasonable request.
% All data used in the article and Supplementary Information are accessible via Zenodo at XXX (ref. XX).

\subsection*{Materials availability} 
Not applicable.

\subsection*{Code availability} 
The simulation and data processing codes that support the plots within this paper will be made available by the corresponding author upon request.

\subsection*{Author contribution} 
D.G.A., P.N.R., J.Z., and C.A.R. conceived the design and planned the experimental methodology. D.G.A. performed the optomechanical simulations. D.G.A. and P.N.R. fabricated the samples with help from S.E. and C.A.R. P.N.R. carried the experimental characterization with the help of P.J.R. and H.E.B.F. P.N.R. performed the numerical calculations for the theoretical model. D.G.A. and P.N.R. analyzed the data. P.C., D.M., E.C., L.V., and D.M.M. contributed to the overall scientific discussions and conceptual framework. N.D.L.K. participated in the project development and provided general scientific input. D.G.A. and C.A.R. secured the funding with help from D.M., E.C., L.V., and D.M.M.. C.A.R. supervised the project. All authors contributed to the manuscript and provided critical feedback.

\bibliographystyle{unsrt}
\bibliography{references}

@article{Noise_dynamics_2016,
    title = {Noise and dynamics in forward Brillouin interactions},
    author = {Kharel, P. and Behunin, R. O. and Renninger, W. H. and Rakich, P. T.},
    journal = {Physical Review A},
    volume = {93},
    issue = {6},
    pages = {063806-063818},
    year = {2016},
    doi = {10.1103/PhysRevA.93.063806},
}

@article{wolf_theoryNLossBS_2015,
    author = {C. Wolff and P. Gutsche and M. J. Steel and B. J. Eggleton and C. G. Poulton},
    journal = {Journal of the Optical Society of America B},
    number = {9},
    pages = {1968 -- 1978},
    title = {Impact of nonlinear loss on stimulated Brillouin scattering},
    volume = {32},
    year = {2015},
    doi = {10.1364/JOSAB.32.001968},
}

@article{wolff_powerlimit35NL_2015,
    title = {Power limits and a figure of merit for stimulated Brillouin scattering in the presence of third and fifth order loss},
    author = {Wolff, Christian and Gutsche, Philipp and Steel, Michael J and Eggleton, Benjamin J and Poulton, Christopher G},
    journal = {Optics express},
    volume = {23},
    number = {20},
    pages = {26628 -- 26638},
    year = {2015},
    doi = {10.1364/OE.23.026628},
}

@article{BS_3toneGain_2021,
    title = {Demonstration of Forward Brillouin Gain in a Hybrid Photonic–Phononic Silicon Waveguide},
    author = {Wang, Kang and Cheng, Ming and Shi, Haotian and Yu, Linfeng and Huang, Chukun and Qin, Senbiao and Zhang, Yi and Kai, Li and Sun, Junqiang},
    journal = {ACS Photonics},
    volume = {8},
    number = {9},
    pages = {2755-2763},
    year = {2021},
    doi = {10.1021/acsphotonics.1c00880},
}

@ARTICLE{OFDR_theory_1993,
  author={Glombitza, U. and Brinkmeyer, E.},
  journal={Journal of Lightwave Technology}, 
  title={Coherent frequency-domain reflectometry for characterization of single-mode integrated-optical waveguides}, 
  year={1993},
  volume={11},
  number={8},
  pages={1377-1384},
  doi={10.1109/50.254098}
}

@article{NLlossSiwg_TPA_Tsang2008,
    title = {Nonlinear optical properties of silicon waveguides},
    author = {H K Tsang and Y Liu},
    journal = {Semiconductor Science and Technology},
    volume = {23},
    number = {6},
    pages = {064007},
    year = {2008},
    doi = {10.1088/0268-1242/23/6/064007},
}

@article{Aeff_Koos2007,
    title = {Nonlinear silicon-on-insulator waveguides for all-optical signal processing},
    author = {C. Koos and L. Jacome and C. Poulton and J. Leuthold and W. Freude},
    journal = {Optics Express},
    volume = {15},
    number = {10},
    pages = {5976--5990},
    year = {2007},
    doi = {10.1364/OE.15.005976},
}

@article{NLlossSiwg_FCA_Dekker2007,
    title = {Ultrafast nonlinear all-optical processes in silicon-on-insulator waveguides},
    author = {R Dekker and N Usechak and M Först and A Driessen},
    journal = {Journal of Physics D: Applied Physics},
    volume = {40},
    number = {14},
    pages = {R249},
    year = {2007},
    doi = {10.1088/0022-3727/40/14/R01},
}

@article{kittlaus_large_2016,
	title = {Large Brillouin amplification in silicon},
	author = {Kittlaus, Eric A. and Shin, Heedeuk and Rakich, Peter T.},
	journal = {Nature Photonics},
	volume = {10},
	number = {7},
	pages = {463--467},
	year = {2016},
	doi = {10.1038/nphoton.2016.112},
}

@article{TutorialBS_APL_Wiederhecker2019,
    author = {Wiederhecker, Gustavo S. and Dainese, Paulo and Mayer Alegre, Thiago P.},
    title = {Brillouin optomechanics in nanophotonic structures},
    journal = {APL Photonics},
    volume = {4},
    number = {7},
    pages = {071101},
    year = {2019},
    doi = {10.1063/1.5088169},
}

@article{laer_BSoptomech_2016,
	title = {Unifying Brillouin scattering and cavity optomechanics},
	volume = {93},
	doi = {10.1103/PhysRevA.93.053828},
	pages = {053828},
	number = {5},
	journal = {Physical Review A},
	author = {{Van Laer}, R. and Baets, R. and {Van Thourhout}, D.},
	year = {2016},
}

@article{van_laer_interaction_2015,
	title = {Interaction between light and highly confined hypersound in a silicon photonic nanowire},
	author = {Van Laer, Rapha{\"e}l and Kuyken, Bart and Van Thourhout, Dries and Baets, Roel},
	journal = {Nature Photonics},
	volume = {9},
	number = {3},
	pages = {199 -- 203},
	year = {2015},
	doi = {10.1038/nphoton.2015.11},
}

@article{van_laer_net_2015,
	title = {Net on-chip Brillouin gain based on suspended silicon nanowires},
	author = {Van Laer, Rapha{\"e}l and Bazin, Alexandre and Kuyken, Bart and Baets, Roel and Van Thourhout, Dries},
	journal = {New Journal of Physics},
	volume = {17},
	number = {11},
	pages = {115005},
	year = {2015},
	doi = {10.1088/1367-2630/17/11/115005},
}

@article{Shin_SiSiN_2013,
   title = {Tailorable stimulated Brillouin scattering in nanoscale silicon waveguides},
   author={Shin, Heedeuk and Qiu, Wenjun and Jarecki, Robert and Cox, Jonathan A. and Olsson, Roy H. and Starbuck, Andrew and Wang, Zheng and Rakich, Peter T.},
   journal = {Nature Communications},
   volume = {4},
   number = {1},
   pages = {1944},
   year = {2013},
   doi = {10.1038/ncomms2943},
}

@article{lei_antiresonant_2024,
  title = {Anti-resonant acoustic waveguides enabled tailorable Brillouin scattering on chip},
  author = {Lei, Peng and Xu, Mingyu and Bai, Yunhui and Chen, Zhangyuan and Xie, Xiaopeng},
  journal = {Nature Communications},
  volume = {15},
  number = {1},
  pages = {3877},
  year = {2024},
  doi = {10.1038/s41467-024-48123-5}
}

@book{boyd_nonlinearOptics_2020,
  title = {Nonlinear Optics (Fourth Edition)},
  author = {Boyd, Robert W.},
  isbn = {9780080479750},
  year = {2020},
  publisher = {Elsevier Science},
  doi = {10.1016/C2015-0-05510-1}
}

@article{Merklein_100yearsBS_2022,
    title = {100 years of Brillouin scattering: Historical and future perspectives},
    author = {Merklein, Moritz and Kabakova, Irina V. and Zarifi, Atiyeh and Eggleton, Benjamin J.},
    journal = {Applied Physics Reviews},
    volume = {9},
    number = {4},
    pages = {041306},
    year = {2022},
    doi = {10.1063/5.0095488},
}

@article{Loh_ultranarrowBSlaser_nKref_2019,
    author = {William Loh and Siva Yegnanarayanan and Frederick O'Donnell and Paul W. Juodawlkis},
    title = {Ultra-narrow linewidth Brillouin laser with nanokelvin temperature self-referencing},
    journal = {Optica},
    volume = {6},
    number = {2},
    pages = {152--159},
    year = {2019},
    doi = {10.1364/OPTICA.6.000152},
}

@article {Soto_BSdistritibutedsensor_2021,
	title = {Evaluating measurement uncertainty in Brillouin distributed optical fibre sensors using image denoising},
	author = {Soto, Marcelo A and Yang, Zhisheng and Ramírez, Jaime A and Zaslawski, Simon and Thévenaz, Luc},
    journal = {Nature Communications},
	volume = {12},
	number = {1},
	pages = {4901},
    year = {2021},
	doi = {10.1038/s41467-021-25114-4},
}

@article{Stiller_nonreciprocalVortexIsolatorBS_2022,
    author = {Xinglin Zeng  and Philip St.J. Russell  and Christian Wolff  and Michael H. Frosz  and Gordon K. L. Wong  and Birgit Stiller },
    title = {Nonreciprocal vortex isolator via topology-selective stimulated Brillouin scattering},
    journal = {Science Advances},
    volume = {8},
    number = {42},
    pages = {eabq6064},
    year = {2022},
    doi = {10.1126/sciadv.abq6064},
}

@article{Feng_NoiseFree_filterBS_2018,
    author = {Cheng Feng and Stefan Preussler and Thomas Schneider},
    title = {Sharp tunable and additional noise-free optical filter based on Brillouin losses},
    journal = {Photonics Research},
    volume = {6},
    number = {2},
    pages = {132--137},
    year = {2018},
    doi = {10.1364/PRJ.6.000132},
}

@article{Gaeta_TunableDelayBSslowlight_2005,
  title = {Tunable All-Optical Delays via Brillouin Slow Light in an Optical Fiber},
  author = {Okawachi, Yoshitomo and Bigelow, Matthew S. and Sharping, Jay E. and Zhu, Zhaoming and Schweinsberg, Aaron and Gauthier, Daniel J. and Boyd, Robert W. and Gaeta, Alexander L.},
  journal = {Physical Review Letters},
  volume = {94},
  number = {15},
  pages = {153902-153906},
  year = {2005},
  doi = {10.1103/PhysRevLett.94.153902},
}

@article{Boyd_storedLight_OpticalFibre_BS_2007,
    author = {Zhaoming Zhu  and Daniel J. Gauthier  and Robert W. Boyd },
    title = {Stored Light in an Optical Fiber via Stimulated Brillouin Scattering},
    journal = {Science},
    volume = {318},
    number = {5857},
    pages = {1748-1750},
    year = {2007},
    doi = {10.1126/science.1149066},
}

@article{eggleton_BrillouinIntegrated_2019,
  title = {Brillouin integrated photonics},
  author = {Eggleton, Benjamin J and Poulton, Christopher G and Rakich, Peter T and Steel, Michael J and Bahl, Gaurav},
  journal = {Nature Photonics},
  volume = {13},
  number = {10},
  pages = {664--677},
  year = {2019},
  doi = {10.1038/s41566-019-0498-z}
}

@article{Rakich_GiantBS_SWG_2012,
    title = {Giant Enhancement of Stimulated Brillouin Scattering in the Subwavelength Limit},
    author = {Rakich, Peter T. and Reinke, Charles and Camacho, Ryan and Davids, Paul and Wang, Zheng},
    journal = {Physical Review X},
    volume = {2},
    number = {1},
    pages = {011008},
    year = {2012},
    doi = {10.1103/PhysRevX.2.011008},
}

@article{Pant_chalcogenideBBS_2011,
    author = {Ravi Pant and Christopher G. Poulton and Duk-Yong Choi and Hannah Mcfarlane and Samuel Hile and Enbang Li and Luc Thevenaz and Barry Luther-Davies and Stephen J. Madden and Benjamin J. Eggleton},
    journal = {Optics Express},
    number = {9},
    pages = {8285 -- 8290},
    title = {On-chip stimulated Brillouin scattering},
    volume = {19},
    year = {2011},
    doi = {10.1364/OE.19.008285},
}

@article{Otterstrom_BSsiLaser_2018,
    author = {Nils T. Otterstrom  and Ryan O. Behunin  and Eric A. Kittlaus  and Zheng Wang  and Peter T. Rakich },
    title = {A silicon Brillouin laser},
    journal = {Science},
    volume = {360},
    number = {6393},
    pages = {1113-1116},
    year = {2018},
    doi = {10.1126/science.aar6113}
}

@article{Marpaung_BS_SiN_2022,
    author = {Roel Botter  and Kaixuan Ye  and Yvan Klaver  and Radius Suryadharma  and Okky Daulay  and Gaojian Liu  and Jasper van den Hoogen  and Lou Kanger  and Peter van der Slot  and Edwin Klein  and Marcel Hoekman  and Chris Roeloffzen  and Yang Liu  and David Marpaung },
    title = {Guided-acoustic stimulated Brillouin scattering in silicon nitride photonic circuits},
    journal = {Science Advances},
    volume = {8},
    number = {40},
    pages = {eabq2196},
    year = {2022},
    doi = {10.1126/sciadv.abq2196}
}

@article{Marpaung_IntegratedBS_tfln_2025,
    author = {Kaixuan Ye  and Hanke Feng  and Randy te Morsche and Chuangchuang Wei  and Yvan Klaver  and Akhileshwar Mishra  and Zheng Zheng  and Akshay Keloth  and Ahmet Tarık Işık  and Zhaoxi Chen  and Cheng Wang  and David Marpaung },
    title = {Integrated Brillouin photonics in thin-film lithium niobate},
    journal = {Science Advances},
    volume = {11},
    number = {18},
    pages = {eadv4022},
    year = {2025},
    doi = {10.1126/sciadv.adv4022},
}

@misc{PNR_SWGPhnC_2024,
    title = {Stimulated Forward Brillouin Scattering in Subwavelength Silicon Membranes},
    author = {Paula {Nuño Ruano} and Jianhao Zhang and David González-Andrade and Daniele Melati and Eric Cassan and Pavel Cheben and Laurent Vivien and Norberto Daniel Lanzillotti-Kimura and Carlos Alonso-Ramos},
    year = {2024},
    eprint = {2402.15403},
    archivePrefix = {arXiv},
    primaryClass = {physics.optics},
}

@article{Dihn_ControlModeConfine_2023,
    author = {Dinh, Thi Thuy Duong and Le Roux, Xavier and Zhang, Jianhao and Montesinos-Ballester, Miguel and Lafforgue, Christian and Benedikovic, Daniel and Cheben, Pavel and Cassan, Eric and Marris-Morini, Delphine and Vivien, Laurent and Alonso-Ramos, Carlos},
    title = {Controlling the Modal Confinement in Silicon Nanophotonic Waveguides through Dual-Metamaterial Engineering},
    journal = {Laser \& Photonics Reviews},
    volume = {17},
    number = {3},
    pages = {2100305},
    year = {2023},
    doi = {10.1002/lpor.202100305},
}

@article{wolff_tutorialBS_2021,
    author = {C. Wolff and M. J. A. Smith and B. Stiller and C. G. Poulton},
    journal = {Journal of the Optical Society of America B},
    number = {4},
    pages = {1243 -- 1269},
    publisher = {Optica Publishing Group},
    title = {Brillouin scattering -- -theory and experiment: tutorial},
    volume = {38},
    year = {2021},
    doi = {10.1364/JOSAB.416747},
}

@article{GonzalezAndrade_SimuTMforBS_2024,
    author = {David Gonz\'{a}lez-Andrade and Paula {Nuño Ruano} and Jianhao Zhang and Eric Cassan and Delphine Marris-Morini and Laurent Vivien and Norberto Daniel Lanzillotti-Kimura and Carlos Alonso-Ramos},
    title = {Enhancing stimulated Brillouin scattering in suspended silicon waveguides through subwavelength nanostructuration [Invited]},
    journal = {Optical Materials Express},
    number = {11},    
    volume = {14},
    pages = {2562--2577},
    year = {2024},
    doi = {10.1364/OME.534474},
}

@article{Halir_lossesSWG_2016,
    title = {Controlling leakage losses in subwavelength grating silicon metamaterial waveguides},
    author = {J. Dar\'{i}o Sarmiento-Merenguel and Alejandro Ortega-Mo\~{n}ux and Jean-Marc F\'{e}d\'{e}li and J. Gonzalo Wang\"{u}emert-P\'{e}rez and Carlos Alonso-Ramos and Elena Dur\'{a}n-Valdeiglesias and Pavel Cheben and \'{I}\~{n}igo Molina-Fern\'{a}ndez and Robert Halir},
    journal = {Optics Letters},
    volume = {41},
    number = {15},
    pages = {3443--3446},
    year = {2016},
    doi = {10.1364/OL.41.003443},
}

@incollection{Wolff_Chapter2_theorySBS_2022,
    author = {Christian Wolff and Christopher G. Poulton and Michael J. Steel and Gustavo Wiederhecker},
    title = {Chapter Two - Theoretical formalisms for stimulated Brillouin scattering},
    editor = {Benjamin J. Eggleton and Michael J. Steel and Christopher G. Poulton},
    series = {Semiconductors and Semimetals},
    publisher = {Elsevier},
    volume = {109},
    pages = {27-91},
    year = {2022},
    booktitle = {Brillouin Scattering Part 1},
    issn = {0080-8784},
    doi = {https://doi.org/10.1016/bs.semsem.2022.04.002},
}

@article{bedoya_notch_filter_pedestral_2015,
    title = {Tunable narrowband microwave photonic filter created by stimulated Brillouin scattering from a silicon nanowire},
    author = {Alvaro Casas-Bedoya and Blair Morrison and Mattia Pagani and David Marpaung and Benjamin J. Eggleton},
    journal = {Opt. Lett.},
    volume = {40},
    number = {17},
    pages = {4154 -- 4157},
    year = {2015},
    doi = {10.1364/OL.40.004154},
}

@article{Gertler_RFfilterSi_2020,
    author = {Gertler, Shai and Kittlaus, Eric A. and Otterstrom, Nils T. and Rakich, Peter T.},
    title = "{Tunable microwave-photonic filtering with high out-of-band rejection in silicon}",
    journal = {APL Photonics},
    volume = {5},
    number = {9},
    pages = {096103},
    year = {2020},
    doi = {10.1063/5.0015174}
}

@article{gertler_MWP_RFfilter_2wg_2022,
    title = {Narrowband microwave-photonic notch filters using Brillouin-based signal transduction in silicon},
    author = {Gertler, Shai and Otterstrom, Nils T and Gehl, Michael and Starbuck, Andrew L and Dallo, Christina M and Pomerene, Andrew T and Trotter, Douglas C and Lentine, Anthony L and Rakich, Peter T},
    journal = {Nature Communications},
    volume = {13},
    number = {1},
    pages = {1947},
    year = {2022},
    doi = {10.1038/s41467-022-29590-0}
}

@article{cheben_subwavelength_2018,
    title = {Subwavelength integrated photonics},
    author = {Cheben, Pavel and Halir, Robert and Schmid, Jens H and Atwater, Harry A and Smith, David R},
    journal = {Nature},
    volume = {560},
    number = {7720},
    pages = {565 -- 572},
    year = {2018},
    doi = {10.1038/s41586-018-0421-7}
}

@article{Cheben_MetamaterialsIntegratedOptics_2023,
    author = {Pavel Cheben and Jens H. Schmid and Robert Halir and Jos\'{e} Manuel Luque-Gonz\'{a}lez and J. Gonzalo Wang\"{u}emert-P\'{e}rez and Daniele Melati and Carlos Alonso-Ramos},
    journal = {Advances in Optics and Photonics},
    number = {4},
    pages = {1033--1105},
    title = {Recent advances in metamaterial integrated photonics},
    volume = {15},
    year = {2023},
    doi = {10.1364/AOP.495828},
}

@misc{comsol61_2023,
	title = {{COMSOL} Multiphysics® Reference Manual v.6.1},
	publisher = {{COMSOL} {AB}},
	date = {2023},
}

@misc{lumerical_2023,
	title = {{ANSYS Lumerical MODE}},
	publisher = {{ANSYS Lumerical Inc.} {AB}},
	date = {2023},
}

@article{Penades_SuspendedSWGMIR_18,
    author = {J. Soler Penad\'{e}s and A. S\'{a}nchez-Postigo and M. Nedeljkovic and A. Ortega-Mo\~{n}ux and J. G. Wang\"{u}emert-P\'{e}rez and Y. Xu and R. Halir and Z. Qu and A. Z. Khokhar and A. Osman and W. Cao and C. G. Littlejohns and P. Cheben and I. Molina-Fern\'{a}ndez and G. Z. Mashanovich},
    title = {Suspended silicon waveguides for long-wave infrared wavelengths},
    journal = {Optics Letters},
    volume = {43},
    number = {4},
    pages = {795--798},
    year = {2018},
    doi = {10.1364/OL.43.000795},
}

@article{Halir_ReviewSWG_anisotropy_2021,
    title = {A review of silicon subwavelength gratings: building break-through devices with anisotropic metamaterials},
    author = {J. M. Luque-González and A. Sánchez-Postigo and A. Hadij-ElHouati and A. Ortega-Moñux and J. G. Wangüemert-Pérez and J. H. Schmid and P. Cheben and I. Molina-Fernández and R. Halir},
    journal = {Nanophotonics},
    volume = {10},
    number = {11},
    pages = {2765--2797},
    year = {2021},
    doi = {doi:10.1515/nanoph-2021-0110},
}

@article{schmidt_suspendedMIR_2019,
  title = {Suspended mid-infrared waveguides for Stimulated Brillouin Scattering},
  author = {Schmidt, M. K. and Poulton, C. G. and Mashanovich, G. Z. and Reed, G. T. and Eggleton, B. J. and Steel, M. J.},
  journal = {Optics express},
  volume = {27},
  number = {4},
  pages = {4976 -- 4989},
  year = {2019},
  doi = {10.1364/OE.27.004976}
}

@book{CRCchemistryPhysics_2014,
    title = {CRC handbook of Chemistry and Physics},
    author = {Haynes, W. M.},
    edition = {95},
    year = {2014},
    publisher = {CRC press},
    isbn = {781482208689}
}

@book{newnham_MechPropertiesMaterials_2005,
  title = {Properties of Materials: Anisotropy, Symmetry, Structure},
  author = {Newnham, R.E.},
  isbn = {9780198520757},
  year = {2005},
  publisher={OUP Oxford}
}

@article{PhotoelaticSi_Biegelsen1974,
  title = {Photoelastic Tensor of Silicon and the Volume Dependence of the Average Gap},
  author = {Biegelsen, D. K.},
  journal = {Physical Review Letters},
  volume = {32},
  issue = {21},
  pages = {1196--1199},
  year = {1974},
  doi = {10.1103/PhysRevLett.32.1196},
}

%TC:endignore
\end{document}

\maketitle

\tableofcontents
\par\noindent\rule{\textwidth}{0.5pt}

\section{Optomechanical simulations}
In intramodal forward stimulated Brillouin scattering (FSBS), co-propagating pump and scattered waves (Stokes and/or anti-Stokes) of identical mode order and polarization couple through a parametrically generated acoustic phonon, satisfying both energy and momentum conservation \cite{TutorialBS_APL_Wiederhecker2019}: $\Omega=\omega_\mathrm{p}-\omega_\mathrm{s}$, $q(\Omega)=k_\mathrm{p}(\omega_\mathrm{p})-k_\mathrm{s}(\omega_\mathrm{s})$, where $\omega$ and $\Omega$ denote the optical and mechanical angular frequencies, and $k(\omega)$ and $q(\Omega)$ their corresponding wavevectors. Efficient FSBS requires that the acoustic excitation is synchronized with the optical beat pattern created by the pump and Stokes fields. In practice, this means that the acoustic phase velocity must be matched to the optical group velocity of the guided mode. When this condition is satisfied, the periodic optical intensity modulation drives a coherent elastic response that propagates with the light, reinforcing the mechanical wave. In intramodal FSBS, the guided acoustic modes have a finite phase velocities, but their energy transport (group velocity) remains extremely slow (near cut-off modes, $q(\Omega) \approx 0$) resulting in prolonged phonon lifetimes and strong resonant coupling to the optical field. In this configuration, the same mechanical mode mediates both the Stokes and the anti-Stokes processes \cite{wolff_tutorialBS_2021}.

The optical and mechanical modes sustained by the structure are calculated using the finite-element method (COMSOL Multiphysics, eigenfrequency study \cite{comsol61_2023}). An isotropic refractive index of $n_\nit{Si} = 3.45$ is considered for the silicon \cite{CRCchemistryPhysics_2014}. For the elastic properties of silicon, an anisotropic medium is considered, with elastic constants $[c_{11}, c_{12}, c_{44}] = [165.64,\allowbreak 63.94,\allowbreak 79.51]$ \unit{\GPa} and photoelastic constants $[p_{11}, p_{12}, p_{44}] = [-0.094,\allowbreak 0.017,\allowbreak -0.051]$ \cite{PhotoelaticSi_Biegelsen1974, newnham_MechPropertiesMaterials_2005, CRCchemistryPhysics_2014}. In the simulations, the waveguides are oriented along the $\vu*{z}$-axis, with the $\vu*{y}$-axis corresponding to the thickness of the waveguide (see Fig. 1a of the main text). The fabricated waveguides are aligned along the crystallographic direction $\langle 110 \rangle$, so a rotation of $\theta=\pi/4$ around the $\vu*{y}$-axis is applied to the elastic and photoelastic tensors of silicon to properly consider the orientation of the waveguides. Finally, a density of $\rho = 2329$ \unit{\kg\per\m^3} is assumed for the silicon \cite{CRCchemistryPhysics_2014}.

To calculate the optomechanical coupling coefficient $G_\nit{B}$, we compute the electrostriction and radiation pressure contributions according to Wiederhecker \textit{et al.} \cite{TutorialBS_APL_Wiederhecker2019}, 
\begin{equation} \label{eq:GB_coefficient}
    G_\nit{B} = \frac{2 \omega_\p}
    {m_{\nit{eff}} \Omega_\m \Gamma_\m} \left|\int f_{\nit{MB}} \nit{d}\ell + \int f_{\nit{PE}} \nit{d}A \right|^2,
\end{equation} 
where $\omega_\p$ is the pump angular frequency, $\Omega_\m, \Gamma_\m$ are the mechanical angular frequency and linewidth respectively,  $m_{\nit{eff}} = \int\rho|\vb{u}|^2 / \max(|\vb{u}|^2)\, \nit{d}A$ is the effective linear mass density of the mechanical mode with displacement profile $\vb{u}$ ($\rho$ is the material density), and $f_{\nit{MB}}$ and $f_{\nit{PE}}$ are the linear and surface overlap integrals representing the moving boundary (MB) and photoelastic (PE) effects, respectively. 

Using a power normalization for the optical and mechanical modes, and assuming the phase-matching condition is fulfilled, the formal expressions for the MB and PE effects are given by
\begin{align} 
    f_\nit{PE} & = \frac{\vb{E}^*_\p \, \delta\e_\nit{PE} \, \vb{E}_\s}{\max(|\vb{u}|) \, P_\p \, P_\s} \label{eq:PE} \\
    f_\nit{MB} & = \frac{ \left(\vb{u}^*\cdot\vb{n} \right) \, \left(\delta\e_\nit{MB} \, \vb{E}^*_\nit{p,t}\cdot \vb{E}_\nit{s,t} - \delta\e_\nit{MB}^{-1} \, \vb{D}_\nit{p,n}^*\cdot\vb{D}_\nit{s,n}\right)}{\max(|\vb{u}|) \, P_\p \, P_\s}.  \label{eq:MB}
\end{align}
The photoelastic effect represents a force volume density per unit of optical power, with dimensions of [\unit{\newton\per\W\per\m\tothe{3}}], and depends on the pump ($j=\p$) and scattered ($j=\s$) electric field $\vb{E}_j$ and the mechanical displacement $\vb{u}$. The moving boundary effect represents a force surface density per unit of optical power, with dimensions of [\unit{\newton\per\W\per\m\tothe{2}}], and depends only on the tangent (t) or normal (n) components of the electric fields and mechanical motion. Additionally, both effects depend on permittivity perturbations as $\delta\e_\nit{PE} = -\e_0 n^4 \vb{p}:\vb{T}$ and $\delta\e_\nit{MB} = \e_1 - \e_2$ and $\delta\e_\nit{MB}^{-1} = \e_1^{-1} - \e_2^{-1}$. Here, $\e_0$ is the permittivity of vacuum, $\e_1$ of the guiding medium, and $\e_2$ of the surrounding medium; $n$ is the guiding material refractive index, $\vb{p}$ the photoelastic tensor, and $\vb{T}$ the mechanical stress tensor induced by the mechanical mode. The optical power is given by $P_j = [\Re\{2\int(\vb{E}_j\times\vb{H}_j) \cdot \vu{z} \,\nit{d}A\}\,]^{1/2}$.

Equations \eqref{eq:GB_coefficient}, \eqref{eq:PE}, and \eqref{eq:MB} apply to translationally invariant waveguides. This is not the case in our acoustic membrane (SWA) waveguide, and thus, we need to consider the 3D geometry to calculate the Brillouin gain coefficient. However, we can still consider the previous expression by taking the average over a single longitudinal period $\Lambda_\mathrm{L}$ of the waveguide, $\int \nit{d}\ell \mapsto 1/\Lambda_\mathrm{L} \int \nit{d}A$ and $\int \nit{d}A \mapsto 1/\Lambda_\mathrm{L} \int \nit{d}V$ \cite{schmidt_suspendedMIR_2019}.

\section{Leakage loss simulations}
Optical leakage losses of SWA membrane  waveguides were evaluated using a finite-difference eigenmode (FDE) solver (ANSYS Lumerical MODE \cite{lumerical_2023}). In suspended waveguides with subwavelength membrane claddings, leakage constitutes an important loss channel and must be carefully quantified \cite{Penades_SuspendedSWGMIR_18}. The periodic nature of subwavelength-grating-suspended waveguides implies that their optical modes are formally described by Floquet-Bloch solutions of Maxwell’s equations \cite{cheben_subwavelength_2018}. In practice, such modes are not readily computed using conventional eigenmode solvers, requiring suitable approximations.

\begin{figure}[H]
    \centering
    \includegraphics[width=\linewidth]{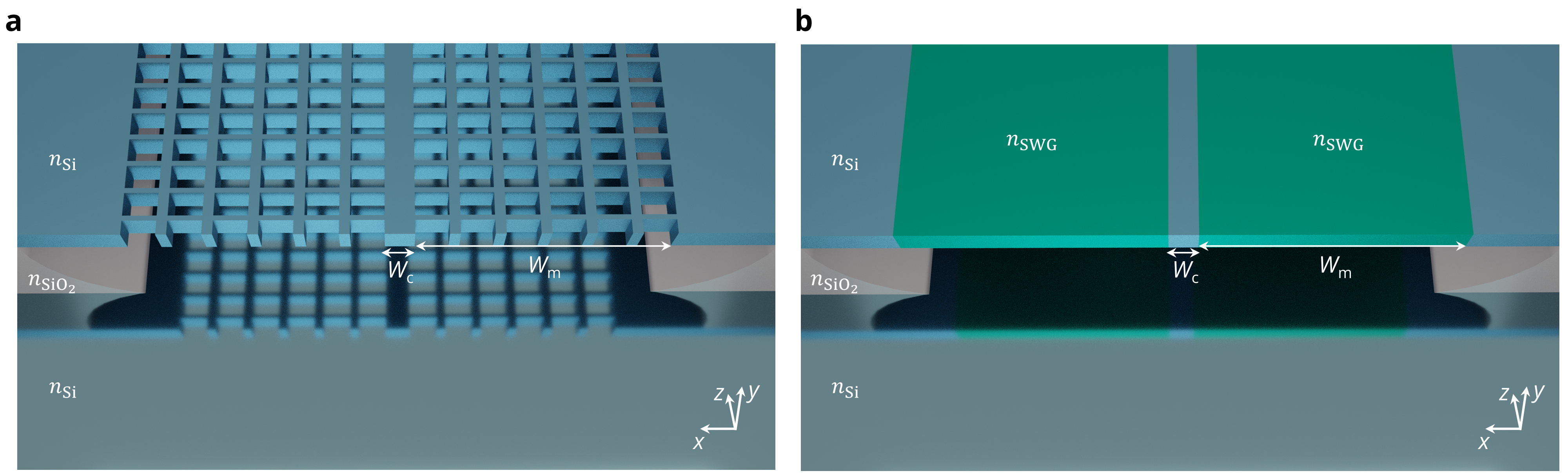}
    \caption{Model of the suspended Brillouin waveguide with lateral metamaterial membranes. \textbf{a} Schematic of the SWA membrane waveguide. \textbf{b} Optical equivalent waveguide in which the lateral subwavelength membranes are modeled as homogeneous dielectric slabs. Simulation parameters: $W_\mathrm{c}$, silicon core width; $W_\mathrm{m}$, membrane width; $n_\mathrm{Si}$, refractive index of silicon; $n_\mathrm{SiO_\mathrm{2}}$, refractive index of silicon dioxide; and $n_\mathrm{SWG}$, effective refractive index of the subwavelength-grating membrane.}
    \label{fig:membrane_model}
\end{figure}

The dual-periodic subwavelength grating (SWG) membrane cladding suppress reflection and diffraction effects and the optical field experiences the cladding as a synthetic dielectric medium (i.e., a metamaterial). Under these conditions, the SWG can be  described by an effective homogeneous material with an equivalent refractive index determined by the silicon fill factor, as shown in Fig. \ref{fig:membrane_model}. This effective-medium approximation is widely used in SWG-based photonic structures and enables efficient and reliable estimation of leakage losses without the need for computationally intensive three-dimensional Floquet-Bloch simulations \cite{Halir_ReviewSWG_anisotropy_2021}.

For each set of core width and membrane periods ($\Lambda_\mathrm{L}$, $\Lambda_\mathrm{T}$), the subwavelength grating (SWG) cladding was replaced by an equivalent homogeneous medium with an effective refractive index $n_\mathrm{SWG}$. The SWG cladding was modelled as an isotropic material such that the effective index of the guided optical mode is preserved relative to the fully periodic structure. This isotropic effective-medium approximation provides an efficient and reliable estimate of substrate leakage without resorting to full three-dimensional Floquet–Bloch simulations, which are computationally more demanding. The resulting waveguide cross-section, including the silicon core, effective-medium SWG cladding, air and silicon substrate, was simulated using an FDE solver with perfectly matched layers (PMLs) to absorb radiated fields at the simulation boundaries. Material absorption was neglected in the simulations, so that the calculated losses reflect only optical leakage.

The FDE solver yields a complex effective index $n_\mathrm{eff}=n_\mathrm{r}+in_\mathrm{i}$ for each guided mode, where the imaginary component captures radiative leakage. The corresponding optical leakage loss $\alpha_\mathrm{leakage}$ (in \unit{\dB\per\m\tothe{1}}) was calculated from the imaginary part of the effective index according to 

\begin{equation}
    \alpha_\mathrm{leakage} = \frac{n_\mathrm{i}40\pi\log_\mathrm{10}(e)}{\lambda_\mathrm{0}},
\end{equation}
with $\lambda_\mathrm{0}$ the free-space wavelength.

By sweeping the waveguide core width and the effective SWG index across the relevant design space, we construct leakage-loss maps of Fig. 2a in the main text, which enable identification of geometries combining strong optical confinement with  low propagation loss.

\section{Optimization of the subwavelength acoustic (SWA) membrane waveguide}
\begin{figure}[H]
    \centering
    \includegraphics[width=\linewidth]{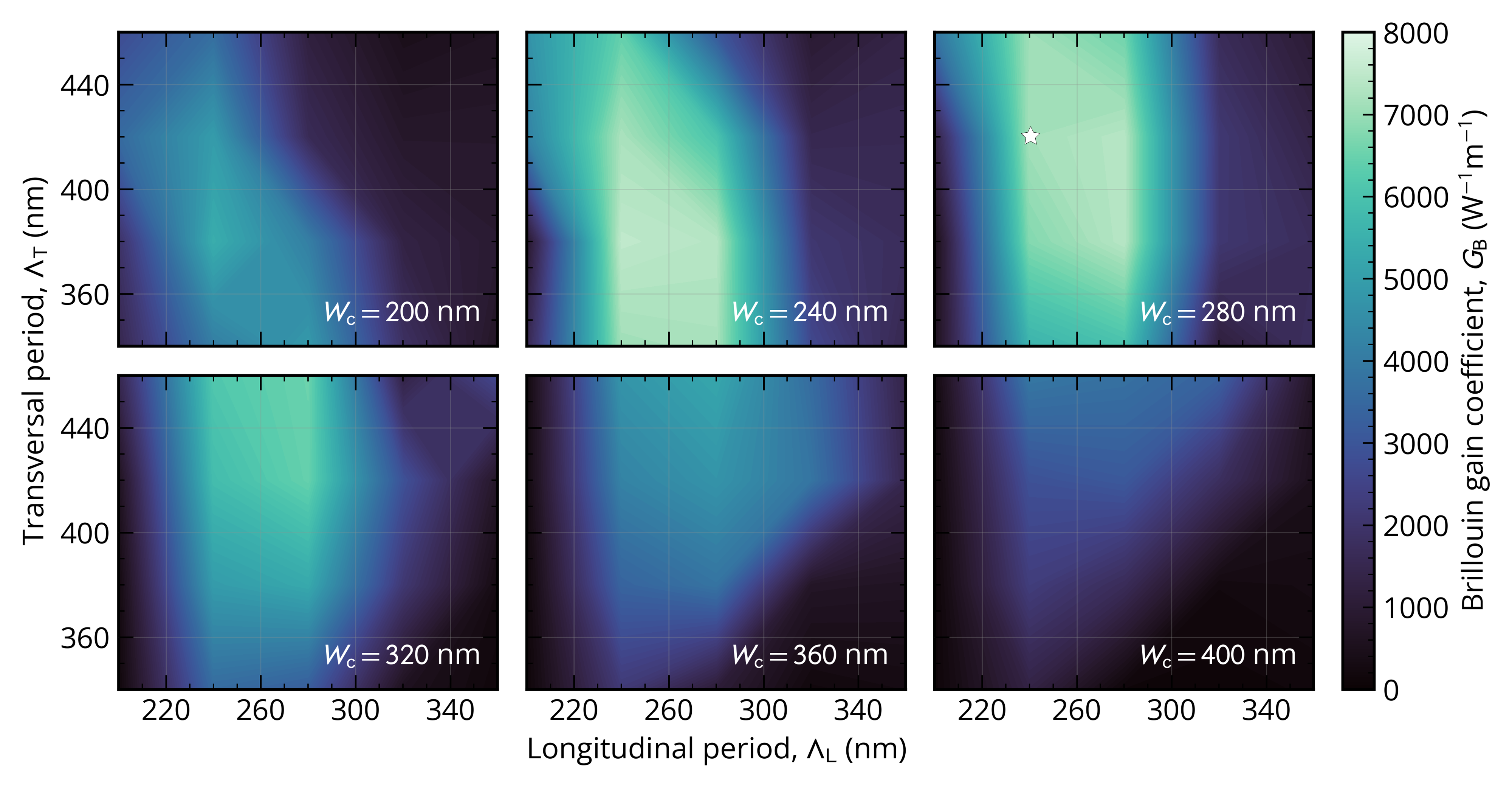}
    \caption{Simulated Brillouin gain coefficient $G_\mathrm{B}$ for different core widths as a function of the membrane periods $\Lambda_\mathrm{L}$ and $\Lambda_\mathrm{T}$.}
    \label{fig:brillouin_gain_all}
\end{figure}

\begin{figure}[H]
    \centering
    \includegraphics[width=\linewidth]{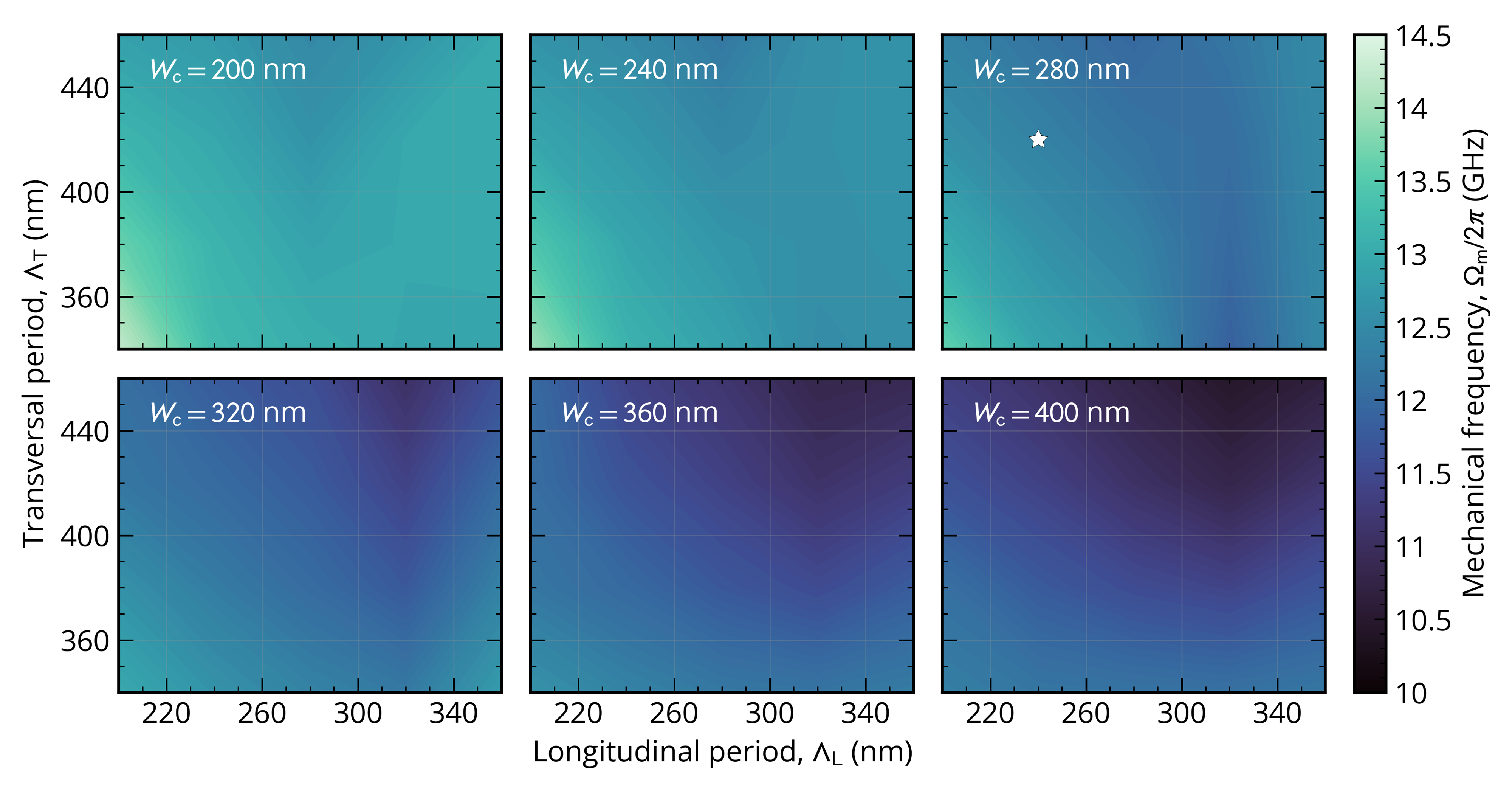}
    \caption{Simulated mechanical frequency $\Omega_\mathrm{m}/2\pi$ for different core widths as a function of the membrane periods $\Lambda_\mathrm{L}$ and $\Lambda_\mathrm{T}$.}
    \label{fig:mechanical_frequency_all}
\end{figure}

Numerical optimization of the SWA membrane waveguide geometry was carried out by evaluating the Brillouin gain coefficient $G_\mathrm{B}$ as a function of the longitudinal and transverse membrane periods ($\Lambda_\mathrm{L}$ and $\Lambda_\mathrm{T}$) for several waveguide core widths $W_\mathrm{c}$. Optomechanical simulations were performed considering a mechanical quality factor fixed at $Q_\mathrm{m}=620$. For each value of $W_\mathrm{c}$, the membrane periods are swept within or near the subwavelength regime. Across all simulated geometries, the highest simulated gain values are obtained for narrow cores, with a maximum occurring for $W_\mathrm{c}=240$ nm (see Fig.~\ref{fig:brillouin_gain_all}). The strong photon-phonon overlap of this configuration yields a $G_\mathrm{B}$ exceeding 7300 \unit{\per\W\per\m} over a broad range of membrane periods, e.g., $\Lambda_\mathrm{L}=[238 - 280]$ nm and $\Lambda_\mathrm{T}=[340 - 414]$ nm. In parallel, we compute the mechanical frequency of the vertically breathing mode as a function of $\Lambda_\mathrm{L}$ and $\Lambda_\mathrm{L}$ for each core width (see Fig.~\ref{fig:mechanical_frequency_all}). The mechanical frequency remains in the  10.48 -- 14.23 GHz range for all simulated geometries, with modest variations induced by changes in the membrane periodicity. In contrast to the Brillouin gain, the mechanical frequency is governed primarily by the dimensions of the waveguide core, in particular its width and thickness. This behavior reflects the fact that the vertically breathing mode is dominated by the effective stiffness and mass of the silicon core, which set the restoring force and inertia of the out-of-plane motion. Variations in the longitudinal and transverse membrane periods mainly affect the surrounding support structure and have less influence on the vertical breathing dynamics.

\begin{figure}[hbtp]
    \centering
    \includegraphics[width=0.5\linewidth]{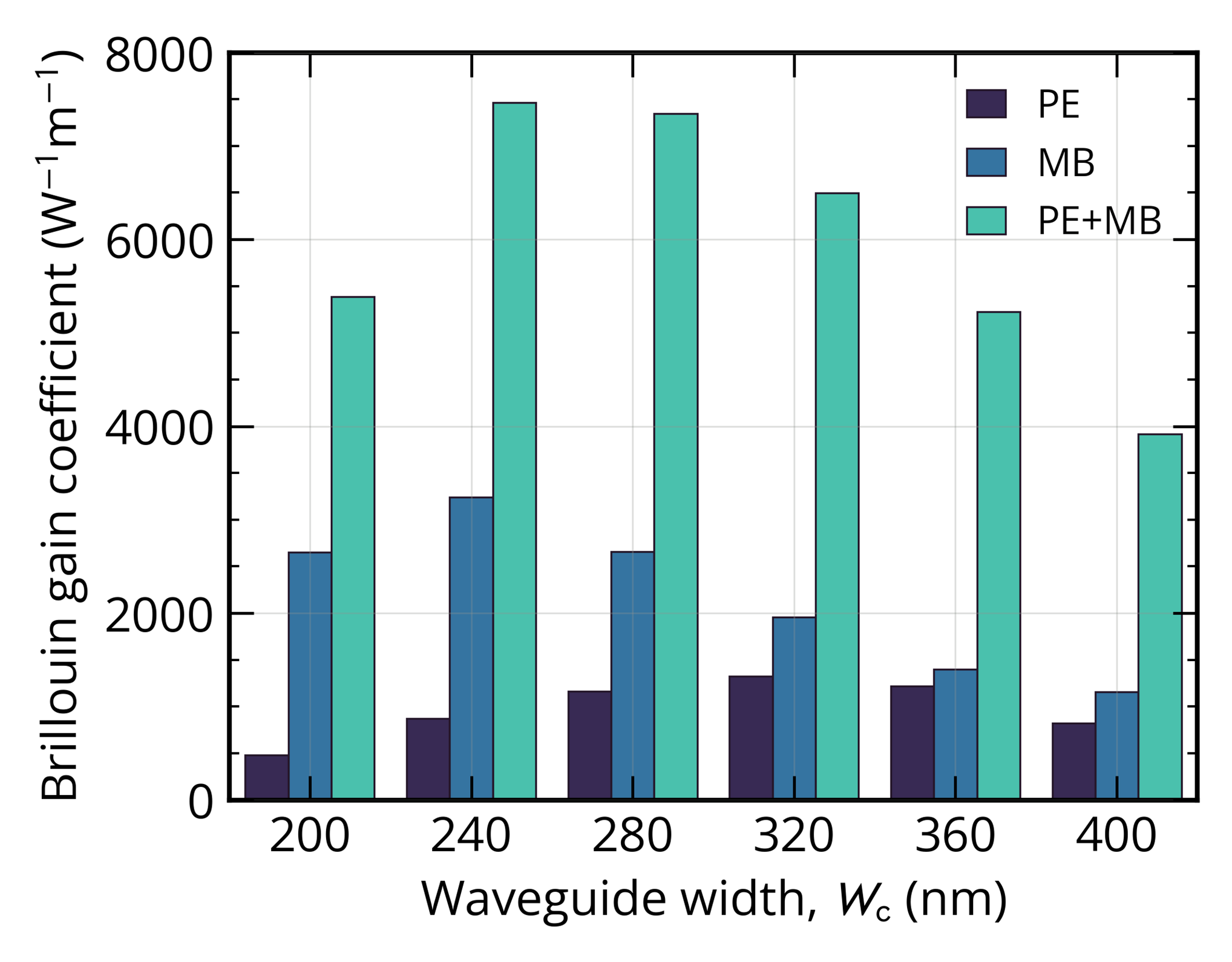}
    \caption{Simulated Brillouin gain coefficient for the optimal pair of membrane periods ($\Lambda_\mathbf{L}$, $\Lambda_\mathbf{T}$) identified for each waveguide core width $W_\mathbf{c}$, together with the corresponding photoelastic (PE) and moving-boundary (MB) contributions.}
    \label{fig:contribution_all}
\end{figure}

Although the largest simulated Brillouin gain is obtained for $W_\mathrm{c}=240$ nm, this core width also exhibits pronounced optical leakage towards the substrate. The narrow core enhances phonon-photon modal overlap but simultaneous increases leakage losses, leading to impractically high propagation loss for experimental implementation. To quantify this trade-off, we perform leakage loss simulations (see Fig. 2a in the main text), identify for each core width the optimal pair of membrane periods ($\Lambda_\mathrm{L}$, $\Lambda_\mathrm{T}$) that maximize $G_\mathrm{B}$, and evaluate the corresponding photoelastic (PE) and moving-boundary (MB) contributions. For all geometries, the MB contribution dominates (see Fig.~\ref{fig:contribution_all}, consistent with the strong vertical displacement of the membrane interfaces and the concentration of the optical field at the top and bottom boundaries. Comparing the designs, the $W_\mathrm{c}=280$ nm configuration emerges as the most favorable compromise, reducing optical leakage by a factor of 30 compared with the $W_\mathrm{c}=240$ nm configuration. While the peak Brillouin gain is slightly reduced compared to the 240 nm core, the decrease is modest, and $G_\mathrm{B}$ remains higher than 7000 \unit{\per\W\per\m}. Importantly, further increasing the core width strongly suppresses optical leakage and lowers propagation loss; however, this comes at the cost of a reduced optomechanical interaction strength, with the Brillouin gain dropping below 4200 \unit{\per\W\per\m} for $W_\mathrm{c}=400$ nm.

Based on this combined analysis of Brillouin gain and optical leakage, we select $W_\mathrm{c}=280$ nm as the geometry for fabrication and experimental investigation. This geometry preserves strong photon–phonon coupling mediated predominantly by moving-boundary effects, while ensuring low-loss optical guidance and mechanical stability over centimeter-scale lengths.

\section{Tolerance to fabrication errors}
\begin{figure}[H]
    \centering
    \includegraphics[width=\linewidth]{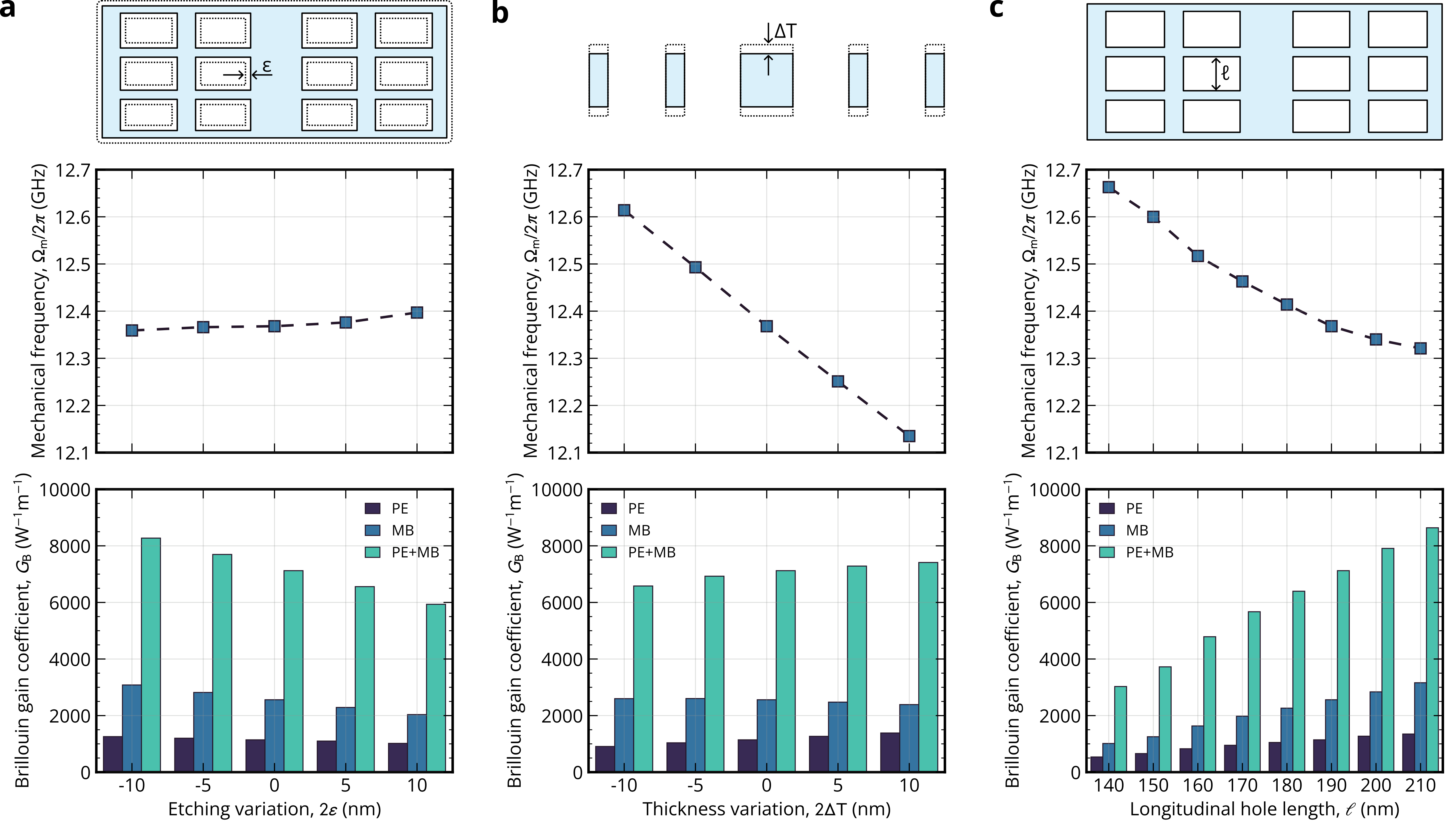}
    \caption{Variations of the nominal SWA membrane waveguide geometry considered in the fabrication tolerance analysis (top panels). Simulated mechanical frequency $\Omega_\mathrm{m}/2\pi$ (mid panels) and Brillouin gain coefficient $G_\mathrm{B}$ (bottom panels) for \textbf{a} etching, \textbf{b} thickness, and \textbf{c} longitudinal hole variations. Nominal geometrical parameters: core width $W_\mathrm{c} = 280$ nm, thickness $T = 300$ nm, and membrane periods $\Lambda_\mathrm{L} = 240$ nm and $\Lambda_\mathrm{T} = 420$ nm.}
    \label{fig:tolerances}
\end{figure}

The robustness of the SWA membrane waveguide design against fabrication imperfections was assessed through simulations performed in COMSOL. Fabrication deviations were introduced by explicitly modifying the device geometry to account for (i) etching errors of $\pm10$ nm, implemented by uniformly offsetting all etched boundaries (i.e., silicon features expand while the air perforations contract, and vice versa); (ii) variations in the silicon device-layer thickness of $\pm10$ nm around the nominal value of 300 nm; and (iii) deviations in the longitudinal hole length, varied from 140 nm to 210 nm while keeping the longitudinal period fixed at $\Lambda_\mathrm{L} = 240$ nm. For each perturbed geometry, the mechanical frequency and the corresponding Brillouin gain coefficient were recalculated (see Fig.~\ref{fig:tolerances}).

The simulations indicate that the mechanical frequency is governed predominantly by the silicon thickness, reflecting the dominant role of vertical confinement in determining the stiffness of the breathing mode. Thickness variations of $\pm$10 nm shift the mechanical frequency from 12.135 to 12.614 GHz, showing that higher (lower) frequencies can be reached by reducing (increasing) the thickness without compromising the overall optomechanical performance. In contrast, both over- and under-etching errors produce only minor changes in mechanical frequency, highlighting the robustness of the vertically breathing mode against overall dimensional fluctuations. When varying the longitudinal hole length alone, smaller holes lead to higher mechanical frequencies, reaching up to 12.663 GHz.

The Brillouin gain coefficient is more sensitive to geometric parameters that govern residual phonon leakage and the effective rigidity of the waveguide core. In particular, reducing the silicon tether dimensions (equivalently increasing the air-hole size) lowers the membrane stiffness while enhancing acoustic confinement, thereby suppressing energy radiation and increasing the Brillouin gain. Variations in the longitudinal hole length at fixed period result in moderate changes in gain, with $G_\mathrm{B} > 6300$ \unit{\per\W\per\m} for typical deviations of $\pm10$ nm around the nominal value of 190 nm. Similarly, thickness variations of $\pm10$ nm maintain Brillouin gain above 6500 \unit{\per\W\per\m}. Across the full range of under- and over-etching errors considered, the Brillouin gain remains above 5900 \unit{\per\W\per\m}, demonstrating substantial tolerance to realistic fabrication errors. 

Overall, these results indicate that the SWA membrane waveguide combines strong photon–phonon coupling with a high degree of fabrication robustness, easing practical implementation.

\section{Nonlinear model for optical losses} \label{sec:NLloss}
The optical power variation along a silicon waveguide can be modeled as \cite{boyd_nonlinearOptics_2020}
\begin{equation}
    \pdv{P}{z} = - \alpha P  - \beta P^2 - \gamma P^3,
\end{equation}
where $\alpha$ stands for the linear power loss coefficient, $\beta$ for the nonlinear power loss coefficient associated with two-photon absorption (TPA), and $\gamma$ for the nonlinear power loss coefficient associated with TPA-induced free-carrier absorption (FCA).

\begin{figure}[H]
    \centering
    \includegraphics[width=0.8\linewidth]{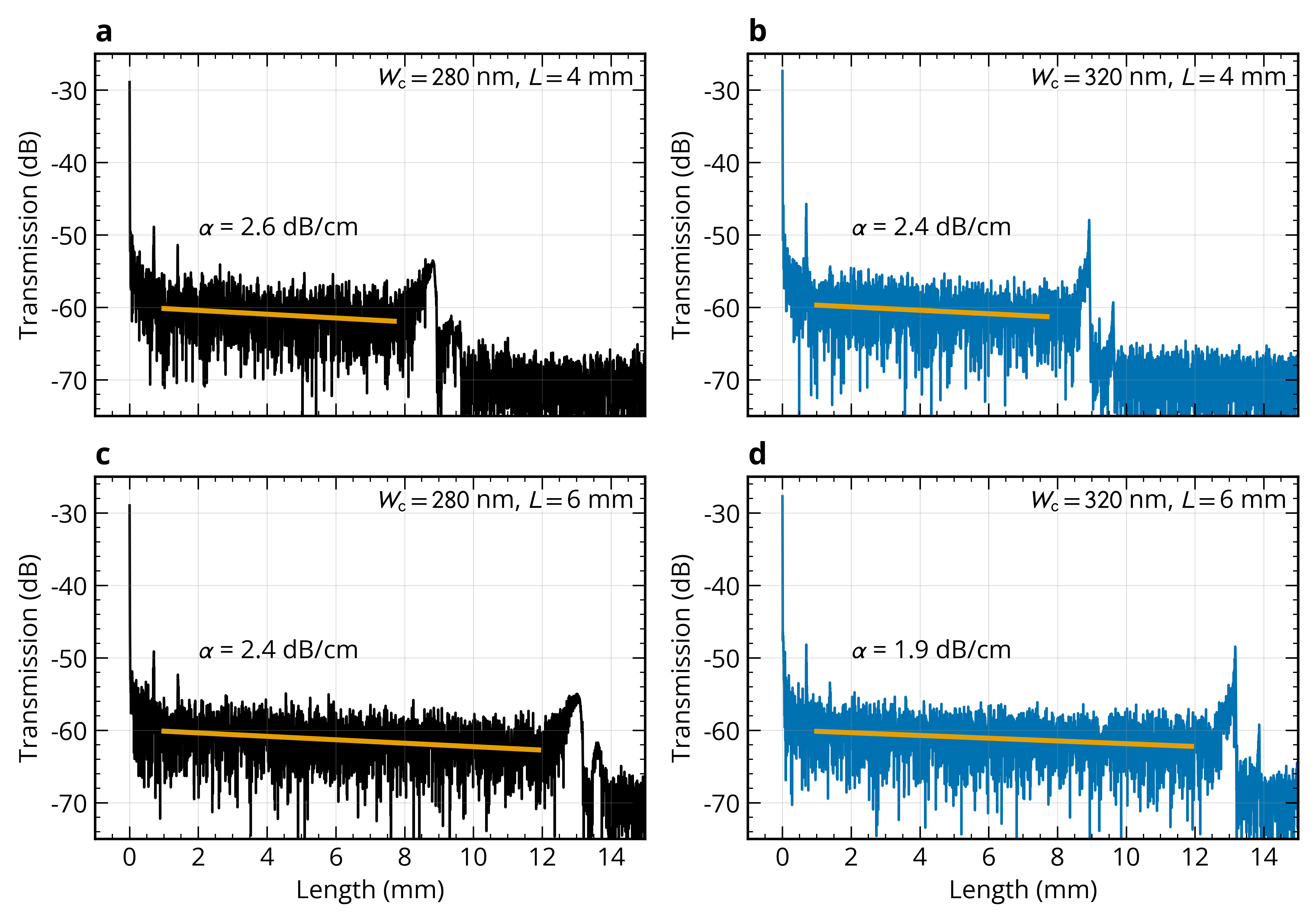}
    \caption{Optical loss for the four waveguides under study measured using the optical frequency domain reflectometry method.}
    \label{fig:optical_loss}
\end{figure}

We determine experimentally the linear loss coefficient, $\alpha$, using optical frequency domain reflectometry (OFDR) \cite{OFDR_theory_1993}, as depicted in Fig. \ref{fig:optical_loss}. The nonlinear loss coefficients are estimated as $\beta = \beta_\mathrm{TPA}/A_\mathrm{eff}$ and $\gamma = \sigma_\mathrm{FCA} \beta_\mathrm{TPA} \tau_\mathrm{c} / (2 h \nu A_\mathrm{eff}^2)$, where $\beta_\mathrm{TPA} \sim 0.6$ \unit{\cm\per\GW} refers to the silicon TPA coefficient \cite{NLlossSiwg_TPA_Tsang2008}, $A_\mathrm{eff}$ is the nonlinear effective mode area for high-index-contrast structures as defined by C. Koos \textit{et al.} \cite{Aeff_Koos2007}, $\sigma_\mathrm{FCA} =$ \num{1.45e-17} \unit{\cm^2} is the free-carrier absorption cross-section \cite{NLlossSiwg_FCA_Dekker2007}, $\tau_\mathrm{c}$ is the free-carrier lifetime, which we assume to be of $\tau_\mathrm{c} = 2.2$ \unit{\ns} \cite{kittlaus_large_2016}, and $h\nu$ is the photon energy ($h\nu = 1.28\cdot 10^{-19}$ J for a 1550 nm wavelength photon). The loss coefficients for the waveguides under study are summarized in Table \ref{tab:loss_coefficients}. 

\begin{table}[H]
    \centering
    \caption{Linear and nonlinear propagation losses for the different waveguides studied in this work.}
    \label{tab:loss_coefficients}
    {\small
    \begin{tabular}{@{}*{8}{c}@{}}
    \toprule
    & && \multicolumn{2}{c}{$W_\nit{c}=280$ nm} && \multicolumn{2}{c}{$W_\nit{c}=320$ nm} \\
    \cmidrule{4-5} \cmidrule{7-8}
    & && 4 mm & 6 mm && 4 mm & 6 mm \\
    \midrule
    $\alpha$ & [\unit{\dB\per\cm}] && 2.6 & 2.4 && 2.4 & 1.9 \\
    $\alpha$ & [\unit{\per\m}] && 59.9 & 55.3 && 55.3 & 43.7 \\
    $\beta$ & [\unit{\per\W\per\m}] && \multicolumn{2}{c}{124} && \multicolumn{2}{c}{121} \\
    $\gamma$ & [\unit{\W^{-2}\m^{-1}}] && \multicolumn{2}{c}{31876} && \multicolumn{2}{c}{30353} \\
    $A_\nit{eff}$ & [\unit{\m^2}] && \multicolumn{2}{c}{$4.84\cdot10^{-14}$} && \multicolumn{2}{c}{$4.96\cdot10^{-14}$} \\
    \bottomrule
    \end{tabular}
    }
\end{table}

\section{Experimental setups} 
Spontaneous intramodal forward Brillouin scattering in low-dispersive waveguides is extremely inefficient \cite{Noise_dynamics_2016}. For this reason, a stimulated regime is needed to obtain a measurable effect. A strong pump and a weak seed for the scattered light are input to the waveguide. Traditionally, the weak seed refers to either the Stokes or anti-Stokes sideband generated by modulating the pump and removing the undesired sideband \cite{van_laer_interaction_2015, kittlaus_large_2016, van_laer_net_2015}. However, this approach requires very narrow and sharp filters. A simpler alternative experimental setup, the so-called three-tone gain experiment \cite{BS_3toneGain_2021}, consists on coupling both Stokes and anti-Stokes sidebands into the waveguide together with the pump. This latter pumping scheme can be understood as a doubly resonant configuration, and has the additional advantage of enhancing the Brillouin on/off response \cite{BS_3toneGain_2021} compared to the former pumping configuration, where only one of the sidebands is present to stimulate the process. In this work, we characterize our waveguide using both configurations to study the full dynamics between pump, Stokes, and anti-Stokes modes, and also provide comparative results with the existing literature.

\subsection{Three-tone pumping scheme}
\begin{figure}[H]
    \centering
    \includegraphics[width=\textwidth]{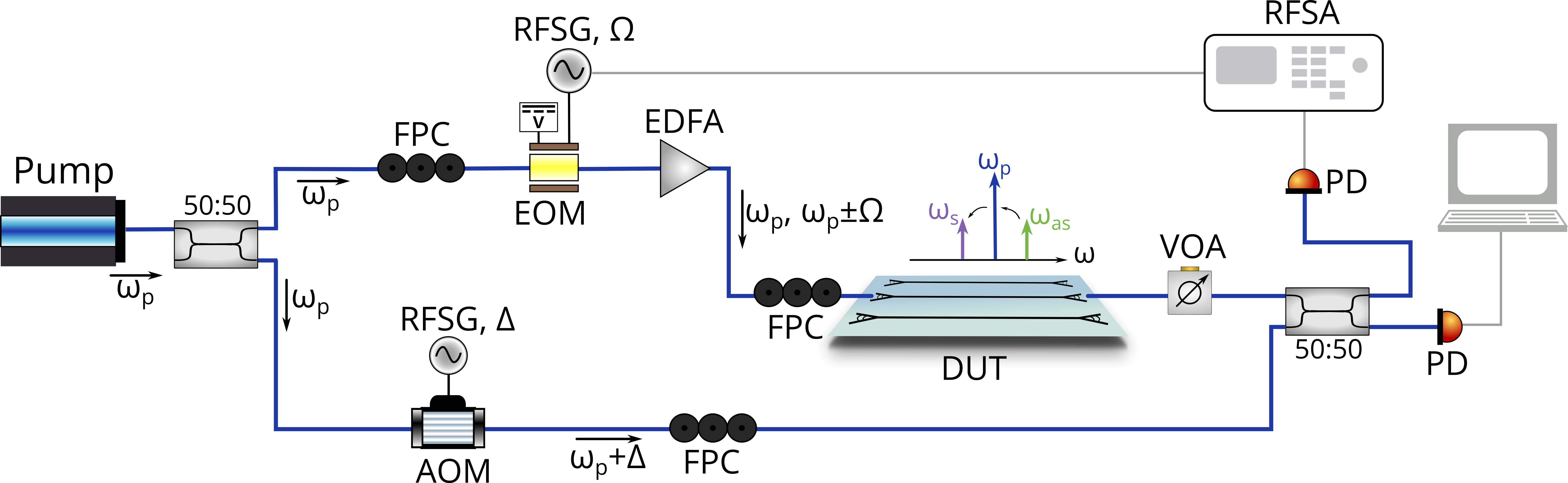}
    \caption{Three-tone gain experimental setup. Two processes of energy transfer take place in the waveguide: from the blue-detuned anti-Stokes sideband $(\omega_\mathrm{as})$ to the pump line $(\omega_\mathrm{p})$, and from the pump line to the red-detuned Stokes sideband $(\omega_\mathrm{s})$. A reference line with up-shifted frequency with respect to the pump $(\omega_\mathrm{p}+\Delta)$ is used to distinguish between both sidelines in the RF spectrum. AOM, acousto-optic modulator; EDFA, erbium-doped fiber amplifier; EOM, electro-optic modulator; FPC, fiber polarization controller; PD, photodetector; RFSA, RF spectrum analyzer; RFSG, RF signal generator;  VOA, variable optical attenuator. In blue, we show the optical links and in grey, the electrical ones.}
    \label{fig:exp-setup-3tone}
\end{figure}

Figure \ref{fig:exp-setup-3tone} depicts the experimental setup for the three-tone gain pumping scheme. Light from a continuous wave (CW) narrow-linewidth laser (Yenista TUNICS TS100-HP) at 1550 nm wavelength is divided into two paths using a 50:50 beam splitter. The light from one of the paths is intensity-modulated using an electro-optic modulator (iXblue MX-LN-40) to generate weak Stokes and anti-Stokes sidebands, and the resulting signal is amplified using an erbium-doped fiber amplifier (EDFA). The operation point for the EOM ensures a difference of $\sim28$ dB between the pump and the sidebands. The three waves are introduced into the device and recovered at the output using focusing grating couplers with an estimated efficiency of $\sim10\%$. A variable optical attenuator is placed at the output to protect the photodetector. The pump light in the second path is up-shifted by 40 MHz using an acousto-optic modulator (AOM, G\&H Fiber-Q 1550 nm 40 MHz) to serve as a reference beam. The light from the chip and the reference beam are combined and then split again using a 50:50 beam combiner. The light is measured by heterodyne detection using a fast photodetector and a radio-frequency (RF) spectrum analyzer (Anritsu MS2830A). Three peaks are observed in the RF domain, corresponding to the frequency difference between the pump and the sidebands at the modulated frequency, $\Omega/2\pi$, the Stokes and the reference at $\Omega/2\pi + 40$ MHz, and the anti-Stokes and the reference at $\Omega/2\pi - 40$ MHz. We sweep the modulation frequency and simultaneously record the RF power of the three beating notes. When the modulation frequency corresponds to the Brillouin shift, a peak in the Stokes line and a dip in the anti-Stokes line are observed simultaneously. All experimental measurements are taken at room temperature and ambient pressure. The resolution bandwidth of the RF spectrum analyzer (RFSA) is set at 10 kHz.

\subsection{Two-tone pump configuration}
\begin{figure}[H]
    \centering
    \includegraphics[width=\linewidth]{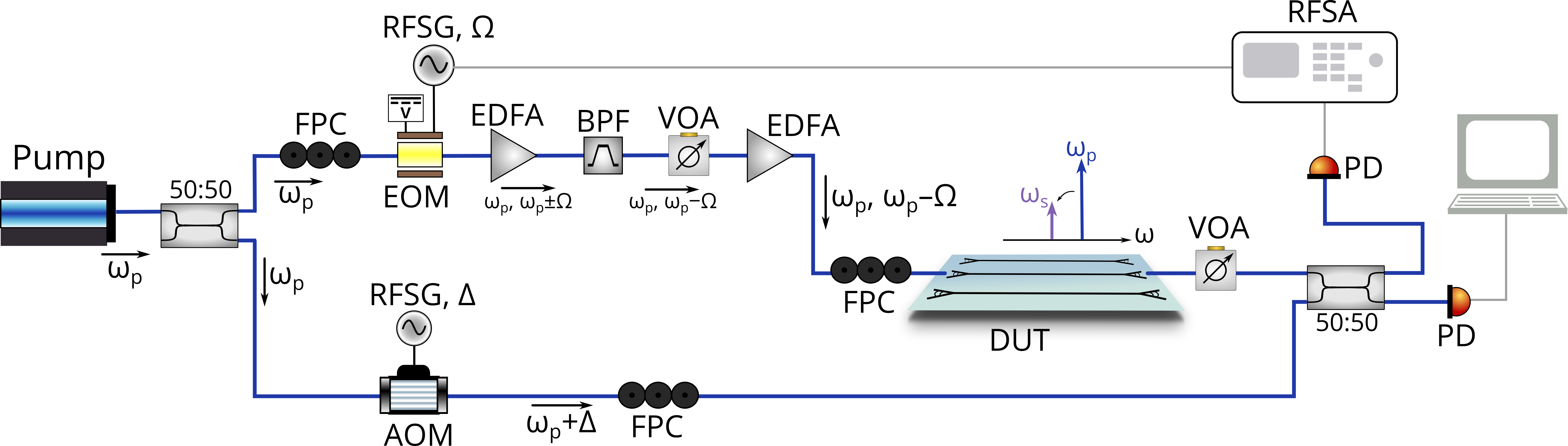}
    \caption{Single-sideband gain experimental setup. As the modulation frequency sweeps across the mechanical resonance, energy is transfer from the pump signal $(\omega_\mathrm{p})$ to the Stokes sideband $(\omega_\mathrm{s})$. A reference line with up-shifted frequency with respect to the pump $(\omega_\mathrm{p}+\Delta)$ is used to distinguish between the sidebands in the RF spectrum to follow their evolution independently. AOM, acousto-optic modulator; BPF, bandpass filter; EDFA, erbium-doped fiber amplifier; EOM, electro-optic modulator; FPC, fiber polarization controller; PD, photodetector; RFSA, RF spectrum analyzer; RFSG, RF signal generator;  VOA, variable optical attenuator. In blue, we show the optical links and in grey the electrical ones.}
    \label{fig:exp-setup-singlesideband}
\end{figure}

Figure \ref{fig:exp-setup-singlesideband} depicts the experimental setup we use to characterize the Brillouin gain using the two-tone strategy. As before, light from a tuneable CW laser (Yenista TUNICS TS100-HP) is intensity-modulated using an electro-optic modulator (iXblue MX-LN-40) to generate two weak sidebands with a power difference of $\sim28$ dB with respect to the pump. A band-pass filter (Yenista XTM-50) removes the undesired sideband and the resulting signal is amplified by an erbium-doped fiber amplifier (EDFA). The additional EDFA and variable optical attenuator (VOA) are used to compensate for the losses in the different elements and ensure optical powers within device requirements. The pump and the sideband seed are introduced in the waveguide and coupled out of it using focusing grating couplers with an efficiency of $\sim10\%$. A reference line obtain by up-shifting the pump frequency by 40 MHz with an acousto-optic modulator (AOM, G\&H Fiber-Q 1550 nm 40 MHz) is added at the output to decouple the sideband and pump dynamics, as well as to calibrate the power difference between both lines. The beating note in the RF domain between the pump, the sideband and the reference is recorded using a fast photodetector and an electrical spectrum analyzer (Anritsu MS2830A) as a function of the modulating frequency.

\section{Theoretical model for intramodal forward Brillouin scattering} \label{sec:3tone_theory}
The dynamics for the intramodal forward Brillouin scattering in waveguides for continuous-wave pump, Stokes, and anti-Stokes signal are described by the following model \cite{Noise_dynamics_2016, wolf_theoryNLossBS_2015, wolff_powerlimit35NL_2015}
\begin{subequations} \label{eq:3tone}
    \begin{align}
    \partial_z A_\p & = -i \frac{G_\mathrm{B}}{2}\left[ \left(\chi|A_\s|^2 + \chi^*|A_\as|^2 \right) \, A_\p + 2 \Re{\chi}A_\s A_\p^* A_\as \right] - \notag\\
     & \mspace{150mu} - \frac{1}{2}\left(\alpha + \beta|A_\p|^2 + \gamma|A_\p|^4\right) A_\p \label{eq:pump_3tone} \\
    \partial_z A_\s & = -i \frac{G_\mathrm{B}}{2} \cdot \chi^* \left(|A_\p|^2 A_\s + A_\as^* A_\p^2 \right) - \frac{1}{2}\left(\alpha + 2\beta|A_\p|^2 + \gamma|A_\p|^4\right) A_\s \label{eq:stokes_3tone} \\
   \partial_z A_\as & = -i \frac{G_\mathrm{B}}{2} \cdot \chi \left(|A_\p|^2 A_\as + A_\s^* A_\p^2 \right)  - \frac{1}{2}\left(\alpha + 2\beta|A_\p|^2 + \gamma|A_\p|^4\right) A_\as \label{eq:antistokes_3tone}
    \end{align}
\end{subequations}
where $A_j$ is the optical mode envelope for the pump ($j=\p$), Stokes ($j=\s$) and anti-Stokes ($j=\as$) fields, normalized to the power, such that $P_j = |A_j|^2$. The term $G_\mathrm{B}$ refers to the Brillouin gain coefficient as defined in equation \eqref{eq:GB_coefficient} \cite{TutorialBS_APL_Wiederhecker2019}, and the mechanical susceptibility $\chi = (\Gamma_\m/2) / [(\Omega-\Omega_\m) + i\Gamma_m/2]$ contains the spectral shape of the interaction. The symbol $^*$ denotes complex conjugate. We include also the linear and nonlinear propagation losses, expressed by the power-related coefficients $\alpha$, $\beta$, and $\gamma$ (see section \ref{sec:NLloss} for a definition of these coefficients). 

Equation \eqref{eq:3tone} is a complex system of coupled differential equations
for the pump, Stokes, and anti-Stokes modes as a result of the same mechanical mode mediating the energy transfer towards the up-shifted and down-shifted lines. Consequently, the power evolution is intrinsically non-exponential, contrary to the conventional backward Brillouin scattering \cite{Noise_dynamics_2016}. Also, the spectral shape for intramodal FBS, given by the mechanical susceptibility, does not corresponds to the typical Lorentzian shape. In order to estimate the relevant parameters (Brillouin gain coefficient, mechanical frequency, and mechanical linewidth), we must first integrate these equations and then, fit the experimental data to the result. Equation \eqref{eq:3tone} is valid for both the three-tone and the two-tone pumping schemes, with the only difference being the initial input powers of the sidebands. 

To gain some insight on the influence of the parameters in the output power, we solve numerically equation \eqref{eq:3tone} for different situations. Unless explicitly indicated otherwise, we consider a three-tone pumping scheme with an input power difference of $-28$ dB between the pump and sidebands. The geometry under study corresponds to a 6 mm-long waveguide with typical experimental values for the optical and mechanical parameters, as detailed in Table \ref{tab:param_simulations_3tone}.

\begin{table}[hbtp]
    \centering
    \caption{Values of the different parameters used for the three-tone gain model simulations.}
    \label{tab:param_simulations_3tone}
    {\small
    \begin{tabular}{@{}*{8}{c}@{}}
    \toprule
    \multicolumn{2}{c}{Geometrical} && \multicolumn{2}{c}{Optical} && \multicolumn{2}{c}{Mechanical} \\
    \midrule
    $W_\nit{c}$ & 280 \unit{\nm} && $\alpha$ & 55.3 \unit{\per\m} && $\Omega_\m/2\pi$ & 12.36 \unit{\GHz}\\
    $L_\nit{wg}$ & 6 \unit{\mm} && $\beta$ & 124 \unit{\per\W\per\m} && $\Gamma_\m/2\pi$ & 20 \unit{\MHz}\\
    \bottomrule
    \end{tabular}
    }
\end{table}

\begin{figure}[H]
    \centering
    \includegraphics[width=\columnwidth]{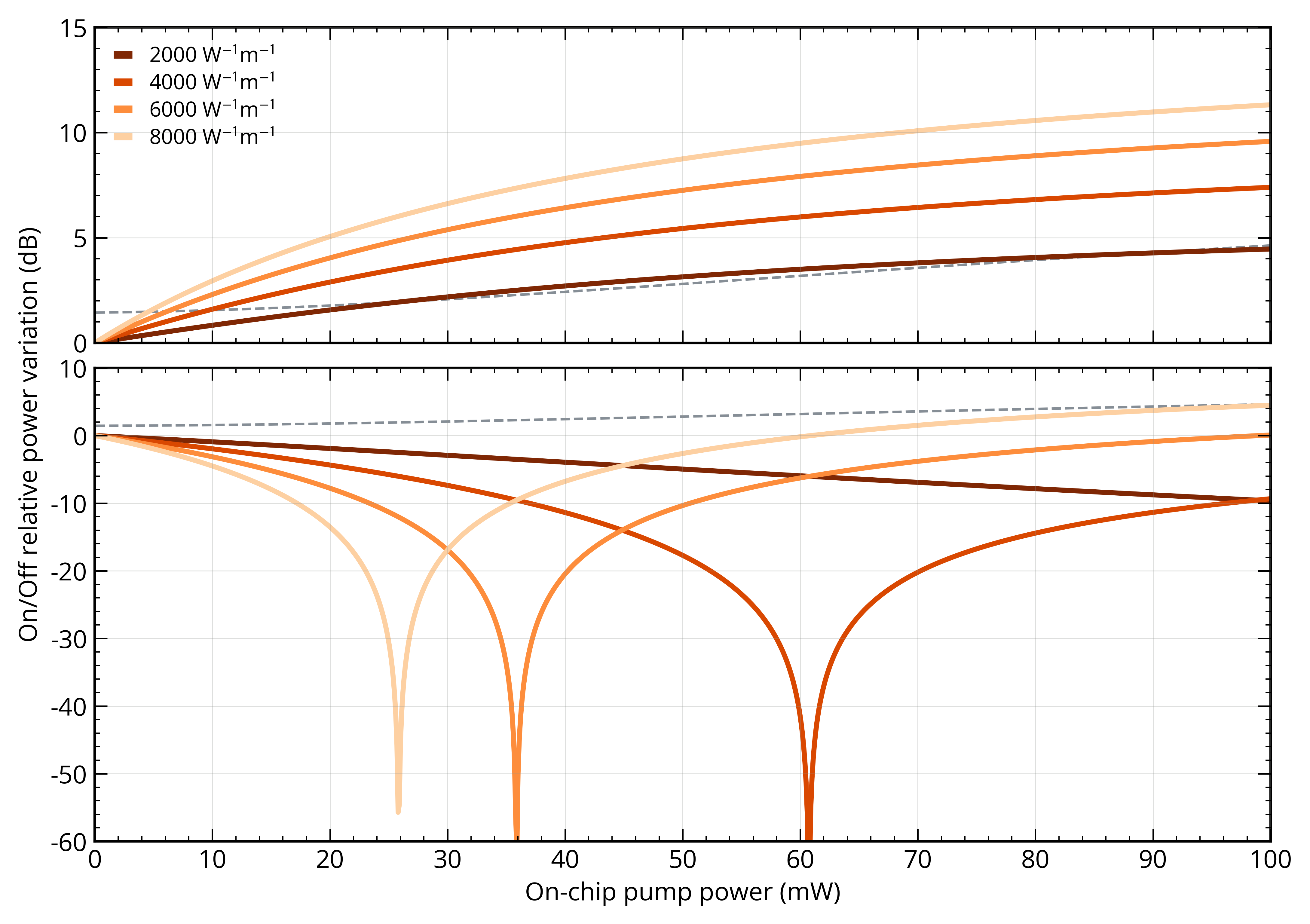}
    \caption{Three-tone pumping scheme. On/off relative power variation for the Stokes (top) and anti-Stokes (bottom) sidebands as a function of the input power for different gain coefficients. The gray dashed line indicates optical propagation losses.}
    \label{fig:3tone_onoff_GB}
\end{figure}

\paragraph{Effect of the gain coefficient in the on/off output signal.} 
Fig. \ref{fig:3tone_onoff_GB} depicts the on/off gain curves for the Stokes (top) and the anti-Stokes (bottom) modes as a function of the input power for different values of the Brillouin gain coefficient. We observe a strong asymmetry in the behavior of each sideband, with stronger depletion than amplification. Additionally, for gain coefficients larger than 3000 \unit{\per\W\per\m}, we observe a change of tendency in the anti-Stokes relative power variation, contrary to the Stokes gain, which monotonically increases with power. The higher the Brillouin gain coefficient, the lower the pump power needed to achieve the maximum depletion (for a fixed waveguide length). Also, for sufficiently high powers and gain coefficients, the model predicts simultaneous amplification for the Stokes and the anti-Stokes modes. In this example, it can be observed for $P_\p>60$ mW and $G_\nit{B}=8000$ \unit{\per\W\per\m}.

\begin{figure}[H]
    \centering
    \includegraphics[width=\linewidth]{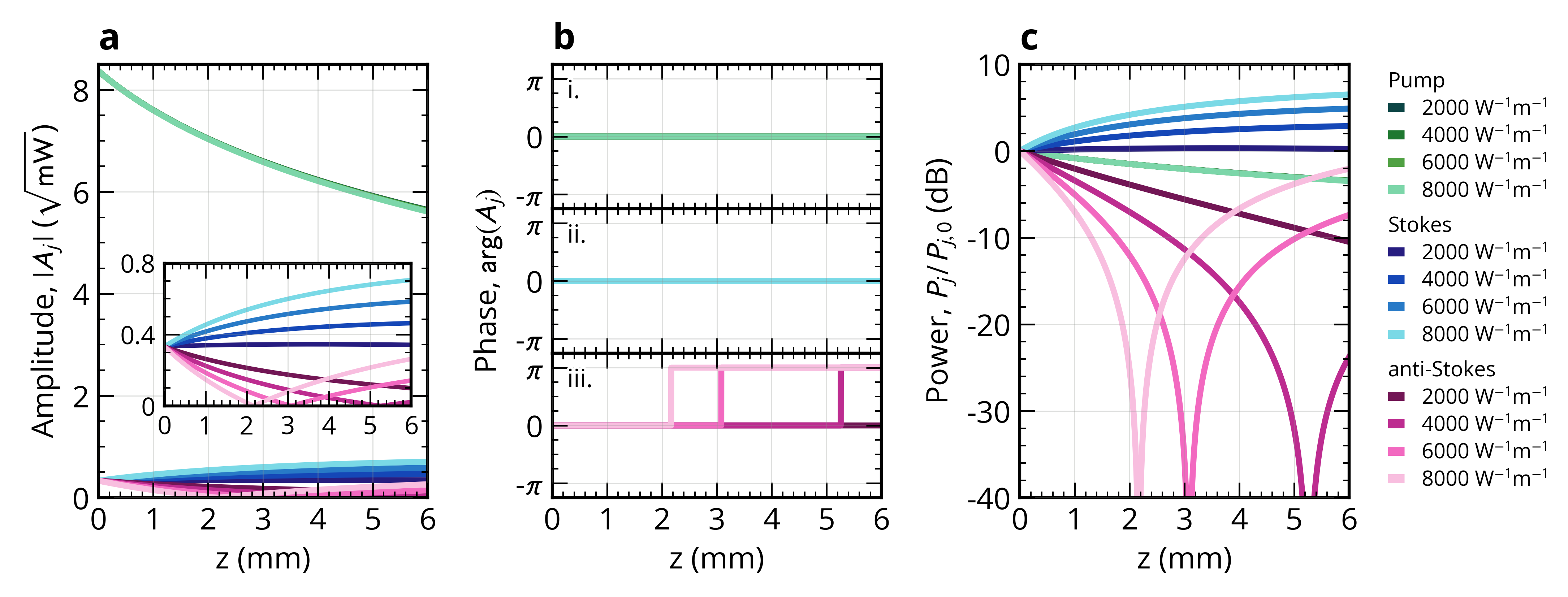}
    \caption{Three-tone pumping scheme. Evolution of the in-resonance amplitude (\textbf{a}), phase (\textbf{b}), and power (\textbf{c}) for the pump (green), Stokes (blue), and anti-Stokes (pink) signals as a function of the waveguide length for different values of the Brillouin gain coefficient. In all cases, a fixed input pump power of 70 mW is assumed, with $-28$ dB of difference for the Stokes and anti-Stokes sidebands.}
    \label{fig:3tone_amplitude_length}
\end{figure}

\paragraph{Evolution of the pump and the sidebands signals along the waveguide.}
To further understand the on/off anti-Stokes depletion curve, we illustrate in Fig. \ref{fig:3tone_amplitude_length} the evolution of the in-resonance pump, Stokes, and anti-Stokes signals along the waveguide length for a fixed input pump power of $P_\p(z=0) = 70$ mW, obtained by integrating equation \ref{eq:3tone}. The pump mode (green lines) is affected mainly by the optical losses, with negligible depletion due to Brillouin scattering, as seen from the similar curves regardless of the gain coefficient. The Stokes sideband (blue lines) shows a monotonic growth of increasing magnitude for larger gain coefficients, and a constant phase. More interesting is the behavior of the anti-Stokes signal (pink lines). For relatively low gain coefficients, the amplitude of this sideband decreases monotonically, and the phase remains constant, in parallel with the Stokes evolution (albeit of a stronger depletion). For large gain coefficients, the anti-Stokes sideband is totally depleted at a given point $z=z_\nit{d}$ of the waveguide, followed by the generation of new anti-Stokes photons with a $\pi$-shift in the phase for the remaining length of the waveguide (Figs. \ref{fig:3tone_amplitude_length}a and \ref{fig:3tone_amplitude_length}b,iii). This behavior appears as a result of the same mechanical mode mediating the Stokes and anti-Stokes process in intramodal forward Brillouin scattering. However, to be observed, an appropriate choice of waveguide length, pump power, and sufficiently large gain coefficient is needed. In log-scale, the total depletion appears a strong dip in the anti-Stokes power (Fig. \ref{fig:3tone_amplitude_length}c).

\begin{figure}[H]
    \centering
    \includegraphics[width=\textwidth]{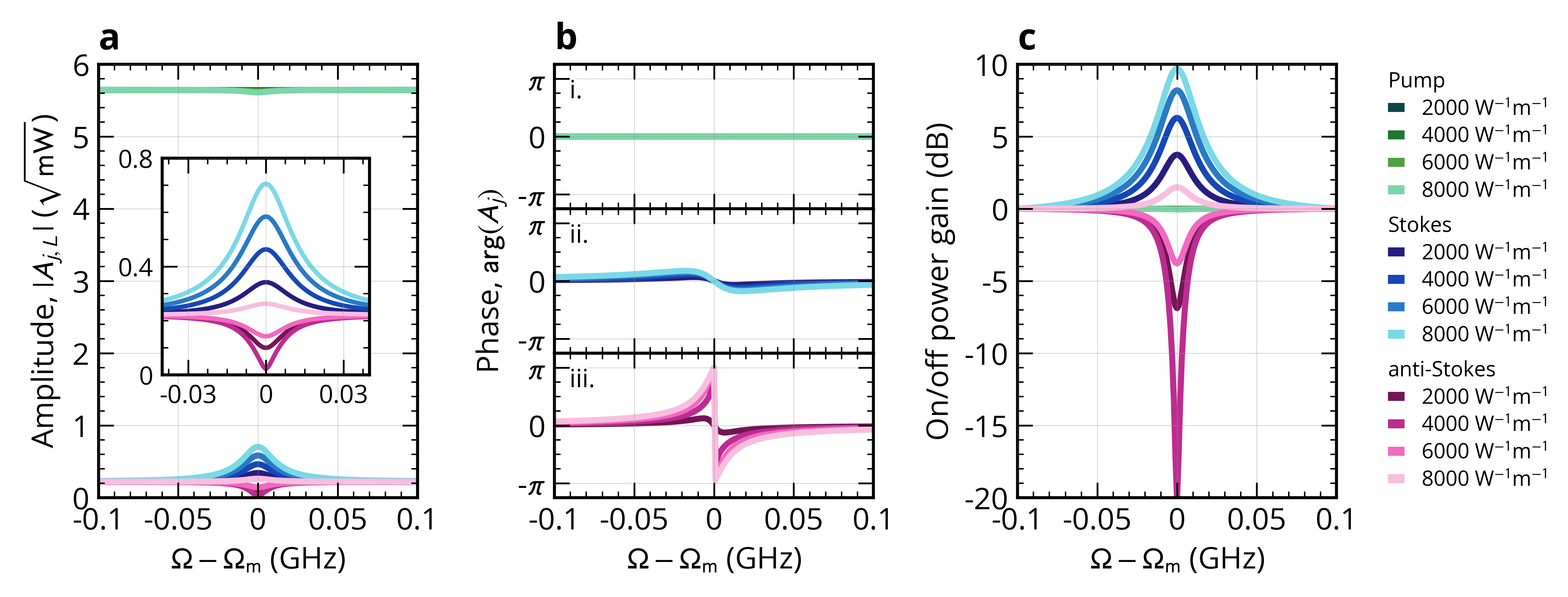}
    \caption{Three-tone pumping scheme.  Evolution of the amplitude (\textbf{a}), phase (\textbf{b}), and power (\textbf{c}) for the pump (green), Stokes (blue), and anti-Stokes (pink) signals as a function of the modulating frequency for different values of the Brillouin gain coefficient. In all cases, a fixed input pump power of 70 mW, with $-28$ dB of difference for the Stokes and anti-Stokes sidebands. The waveguide is assumed to be 6 mm-long.}
    \label{fig:3tone_amplitude_freq}
\end{figure}

\paragraph{Evolution of the output pump and sidebands signals with the modulating frequency.} 
In Fig. \ref{fig:3tone_amplitude_freq}, we perform a similar analysis of the pump, Stokes, and anti-Stokes as a function of the modulating frequency for different values of the Brillouin gain coefficient. We consider a 6 mm-long waveguide and 70 mW of input pump power. This analysis, in particular the on/off gain (Fig. \ref{fig:3tone_amplitude_freq}c) corresponds to the type of experimental measurements detailed in the main text. We observe a negligible effect on the pump power (green lines) as we sweep across the mechanical resonance. The Stokes sideband (blue lines) shows a peak centered at the mechanical resonance, with increasing amplitude for larger gain coefficients. This amplitude peak is accompanied by a small and smooth phase change when crossing the resonance. As in the previous analysis, the most intriguing behavior is displayed by the anti-Stokes signal (pink lines). For low and moderate values of the gain coefficient, the amplitude and power of the anti-Stokes sideband presents a clear dip centered at the mechanical frequency. However, the depth of this dip is not directly proportional to the  $G_\nit{B}$ coefficient. For a fixed length and pump power, it exists a given value for the gain coefficient (here, $\sim4000$ \unit{\per\W\per\m}) that depletes completely the anti-Stokes line when on resonance at the output of the waveguide. For lower $G_\nit{B}$ values, the optomechanical interaction cannot deplete entirely the anti-Stokes line before the end of the waveguide. For larger values, the anti-Stokes depletes fully at some point in the waveguide and starts growing again. As a result, we observe a weaker depletion dip. If the $G_\nit{B}$ increases above a certain value (here, we show the example of 8000 \unit{\per\W\per\m}), the anti-Stokes line displays a gain peak at the mechanical frequency similar to the case of the Stokes sideband. For such values, both Stokes and anti-Stokes modes experience gain simultaneously. To observe this effect experimentally, we need an appropriate combination of pump power, waveguide length, and sufficiently high gain coefficient. The different output behavior is also observed in the phase of the anti-Stokes line (Fig. \ref{fig:3tone_amplitude_freq}b, iii.). For low gain coefficients, the change of phase follows a smooth curve around the mechanical frequency similar to the Stokes line. However, for large $G_\nit{B}$ values, the anti-Stokes sideband presents an abrupt change of phase at the mechanical frequency. 

\subsection{Undepleted pump and small-signal approximations.}
In most experimental configuration, the Stokes and anti-Stokes amplitudes are much smaller than the pump ($|A_\p| \gg |A_\s|,\, |A_\as|$) for the full length of the waveguide, and the optomechanical interaction in the pump signal is negligible, as seen in Fig. \ref{fig:3tone_amplitude_length}.
Thus, we can neglect all the terms that depend quadratically on the sidebands' amplitudes in equation \eqref{eq:pump_3tone}, which decouples equation \eqref{eq:pump_3tone} from equations \eqref{eq:stokes_3tone} and \eqref{eq:antistokes_3tone}. Also, since the remaining term in the right-hand side of equation \eqref{eq:pump_3tone} is real-valued, the pump phase stays constant along the propagation. If we choose for simplicity an initial real-valued pump amplitude $A_\p = \sqrt{P_\p}$, we can rewrite equations \eqref{eq:stokes_3tone} and \eqref{eq:antistokes_3tone} in terms of the pump power, which can be solved independently:
\begin{subequations}
    \begin{align}
    \partial_z P_\p & \approx - \left(\alpha + \beta P_\p + \gamma P_\p^2\right) P_\p \\
    \partial_z A_\s & \approx -i \frac{G_\mathrm{B}}{2} \cdot \chi^* \left(A_\s + A_\as^*\right) P_\p - \frac{1}{2}\left(\alpha + 2\beta  P_\p + \gamma  P_\p^2\right) A_\s \\
   \partial_z A_\as & \approx -i \frac{G_\mathrm{B}}{2} \cdot \chi \left(A_\s^* + A_\as\right) P_\p  - \frac{1}{2}\left(\alpha + 2\beta  P_\p + \gamma  P_\p^2\right) A_\as 
    \end{align}
\end{subequations}
Remembering that $P_j = |A_j|^2 = A_j A_j^*$ ($j=\nit{p,s,as}$), and $\partial_zP_j = A_j (\partial_zA_j)^* + A_j^*\partial_zA_j$, we can introduce the optical power,
\begin{subequations}\label{eq:model_small_signal}
    \begin{align}
    \partial_z P_\p & \approx - \left(\alpha + \beta P_\p + \gamma P_\p^2 \right) P_\p, \label{eq:ss_pump} \\
    \partial_z P_\s & \approx G_\nit{B}\mathcal{L}(\Omega) P_\s P_\p - G_\nit{B} \Im{\chi A_\s A_\as} P_\p^2 - \left(\alpha + 2\beta P_\p + \gamma P_\p^2 \right) P_\s,  \label{eq:ss_stokes} \\
   \partial_z P_\as & \approx - G_\nit{B}\mathcal{L}(\Omega) P_\as P_\p - G_\nit{B} \Im{\chi^* A_\s A_\as} P_\p^2 - \left(\alpha + 2\beta P_\p + \gamma P_\p^2 \right) P_\as, \label{eq:ss_antistokes}
\end{align}
\end{subequations}
where the Lorentzian shape $\mathcal{L}(\Omega)=(\Gamma_\m/2)^2/[(\Gamma_\m/2)^2 + (\Omega-\Omega_\m)^2]$ relates to the mechanical susceptibility as $\mathcal{L}(\Omega)=-\Im{\chi}$. Equation \eqref{eq:model_small_signal} is valid for both the three-tone and the two-tone pumping schemes, with the only difference being the initial input powers of the sidebands. 

Most of the FBS experiments to date \cite{van_laer_net_2015, van_laer_interaction_2015, kittlaus_large_2016} implement a two-tone pumping scheme and the results are analyzed using the so-called \textit{small-signal} limit. This model considers a single sideband to stimulate the Brillouin interaction and implicitly assumes the second sideband remains close to zero for the full length of the waveguide \cite{Noise_dynamics_2016, wolf_theoryNLossBS_2015, wolff_powerlimit35NL_2015}. Therefore, it neglects the coupling term between the Stokes and anti-Stokes term ($\propto A_\s A_\as$) in equation \eqref{eq:model_small_signal}. However, we show below that this coupling term cannot be ignored for moderate powers, large gain coefficients, and/or long waveguides \cite{Wolff_Chapter2_theorySBS_2022}.  

\begin{figure}[H]
    \centering
    \includegraphics[width=\textwidth]{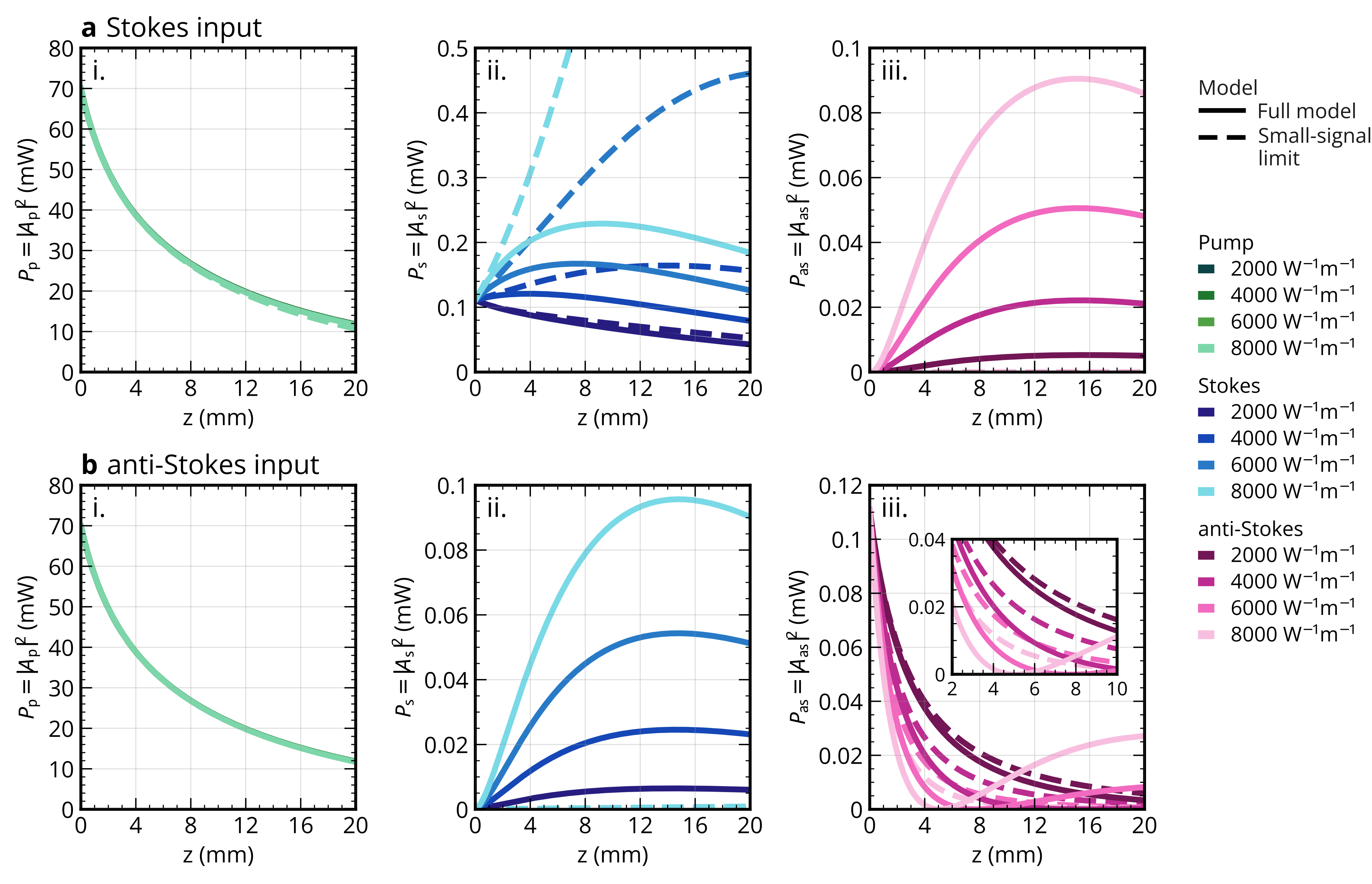}
    \caption{Small-signal limit. Comparison between the full theoretical model and the small-signal limit when \textbf{a} only the Stokes and \textbf{b} only the anti-Stokes is coupled into the waveguide. In each, we plot the evolution of pump power (green, i), Stokes power (blue, ii), and anti-Stokes power (pink, iii) as a function of the position in the waveguide for different values of the Brillouin gain coefficient. In all cases, a fixed input pump power of 70 mW, with $-28$ dB of difference for the Stokes or anti-Stokes sidebands. The solid lines represent the full theoretical model and the dashed lines represents the small-signal limit.}
    \label{fig:3tone_smallsignal_comparison}
\end{figure}

In Fig. \ref{fig:3tone_smallsignal_comparison} we plot the results of the simulated evolution for the pump (left column, green lines), the Stokes (middle column, blue lines), and the anti-Stokes (right column, pink lines) power as a function of the length position in the waveguide for different values of the Brillouin gain coefficient and a fixed input power of 70 mW. We start looking at the case when the anti-Stokes sideband is filtered out (Fig. \ref{fig:3tone_smallsignal_comparison}a). The pump power decreases exponentially along the waveguide, with negligible differences between both models. Also, the pump power evolution is independent on the Brillouin gain coefficient, validating the undepleted pump approximation. The Stokes power increases considerably less if the full model (equation \eqref{eq:3tone}, solid lines) is considered compared to the predictions from the small-signal limit (equation \eqref{eq:model_small_signal}, dashed lines). Only in the case of small Brillouin gain coefficient (in this example, $G_\nit{B} = 2000$ \unit{\per\W\per\m}) the two models give similar results. As for the anti-Stokes power, the small-signal limit predicts zero power for a zero-input anti-Stokes. However, when using the full model, we observe a non-negligible increase in the anti-Stokes power. For the case where the Stokes sideband is the one removed at the input (Fig. \ref{fig:3tone_smallsignal_comparison}b), the pump power evolves as in the previous case, with no differences in the predictions of each model. The Stokes power evolution is the symmetric case of the anti-Stokes before: the small-signal limit shows a constant zero-power while the full model results predicts the generation of Stokes phonons from the noise. The small-signal anti-Stokes power decreases exponentially along the waveguide, reaching asymptotically the zero-power for long waveguides. The full model, nevertheless, predicts the full deplection of anti-Stokes at some point in the waveguide with a faster rate of anti-Stokes photons annihilation that the small-signal time rate. After that, the model predicts the creation of new anti-Stokes photons from the noise, with non-negligible power values if the Brillouin gain is sufficiently large (in this example, $G_\nit{B} = 8000$ \unit{\per\W\per\m}).

\section{Other experimental measurements}
\begin{figure}[H]
    \centering
    \includegraphics[width=\textwidth]{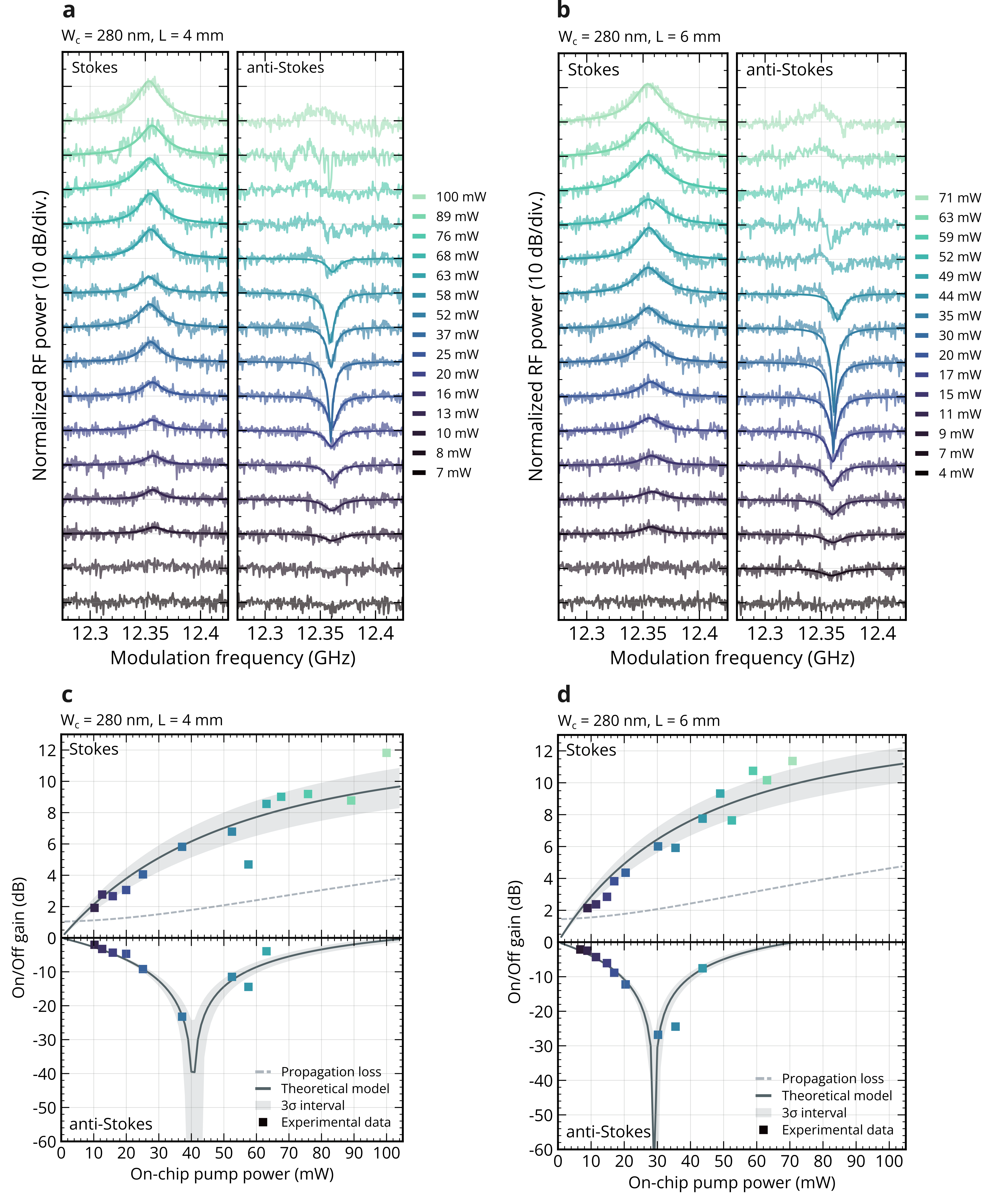}
    \caption{Three-tone pump experiment for the waveguide with core width of $W_\nit{c}=280$ nm. \textbf{a}, \textbf{b} Normalized RF beat note between the reference and the Stokes (left panel) and anti-Stokes (right panel) lines as a function of the modulating frequency for different input pump powers and different lengths (4 mm and 6 mm). \textbf{c}, \textbf{d} On/off gain for the Stokes (upper panel) and anti-Stokes (lower panel) modes at the output of the waveguide as a function of the input pump power and different lengths (4 mm and 6 mm). Figures \textbf{b} and \textbf{d} appear in the main text.}
    \label{fig:3tone_extra_measurements_w280}
\end{figure}

\begin{figure}[H]
    \centering
    \includegraphics[width=\textwidth]{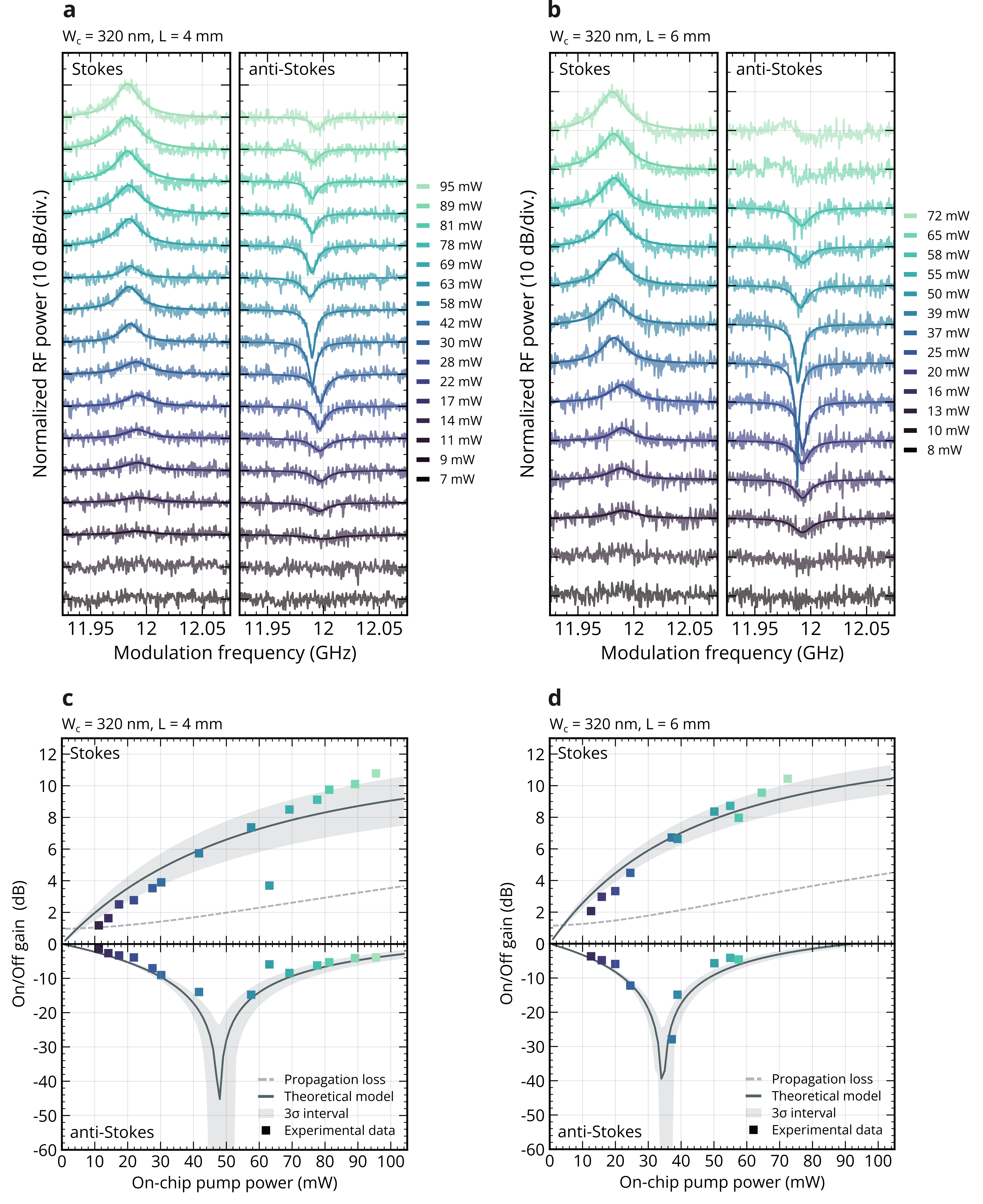}
    \caption{Three-tone pump experiment for the waveguide with core width of $W_\nit{c}=320$ nm. \textbf{a}, \textbf{b} Normalized RF beat note between the reference and the Stokes (left panel) and anti-Stokes (right panel) lines as a function of the modulating frequency for different input pump powers and different lengths (4 mm and 6 mm). \textbf{c}, \textbf{d} On/off gain for the Stokes (upper panel) and anti-Stokes (lower panel) modes at the output of the waveguide as a function of the input pump power and different lengths (4 mm and 6 mm)}
    \label{fig:3tone_extra_measurements_w320}
\end{figure}

We measure the Brillouin scattering of two different families of SWA membrane waveguides for this study, with the same SWG cladding and two different core widths: $W_\nit{c} = 280$ nm and $W_\nit{c} = 320$ nm. In all cases, we normalize the power measurements to the spectra taken at low input power to remove the background associated with the modulator and the detector response. In Figs. \ref{fig:3tone_extra_measurements_w280} and \ref{fig:3tone_extra_measurements_w320}, we illustrate the three-tone pump characterization of two different waveguide lengths, 4 mm and 6 mm, for each of these families. In the four waveguides, we observed a monotonic growth of the Stokes gain as a function of the power, while the anti-Stokes loss increases up to a maximum value and then reduces. The pump power at which the total depletion of the anti-Stokes occurs at the output of the waveguide, characterized by a strong dip in the on/off gain curve, depends on the length and the Brillouin gain coefficient: the longer the waveguide and larger Brillouin gain coefficient, the smaller this \textit{depletion power}. Additionally, as explained in the main text and in section \ref{sec:3tone_theory} (Fig. \ref{fig:3tone_amplitude_freq}), in some of the waveguides we observe an amplification peak for the anti-Stokes line when pumping with sufficiently power. This regime has not been previously observed in integrated waveguides.

In Table \ref{tab:3tone_results}, we summarize the main optomechanical parameters obtained by fitting the experimental data to the theoretical model described in section \ref{sec:3tone_theory}. Overall, a wider core width yields a lower Brillouin gain coefficient due to a lower moving-boundaries effect, as shown in Fig. \ref{fig:contribution_all}. The associated mechanical mode has a lower frequency, as expected from the simulations (see Fig. \ref{fig:mechanical_frequency_all}). The mechanical quality factor is comparable between the different waveguides under study, as the mechanical losses are mainly determined by the SWG cladding, whose design is the same in all the devices we measured.

\begin{figure}[H]
    \centering
    \includegraphics[width=\textwidth]{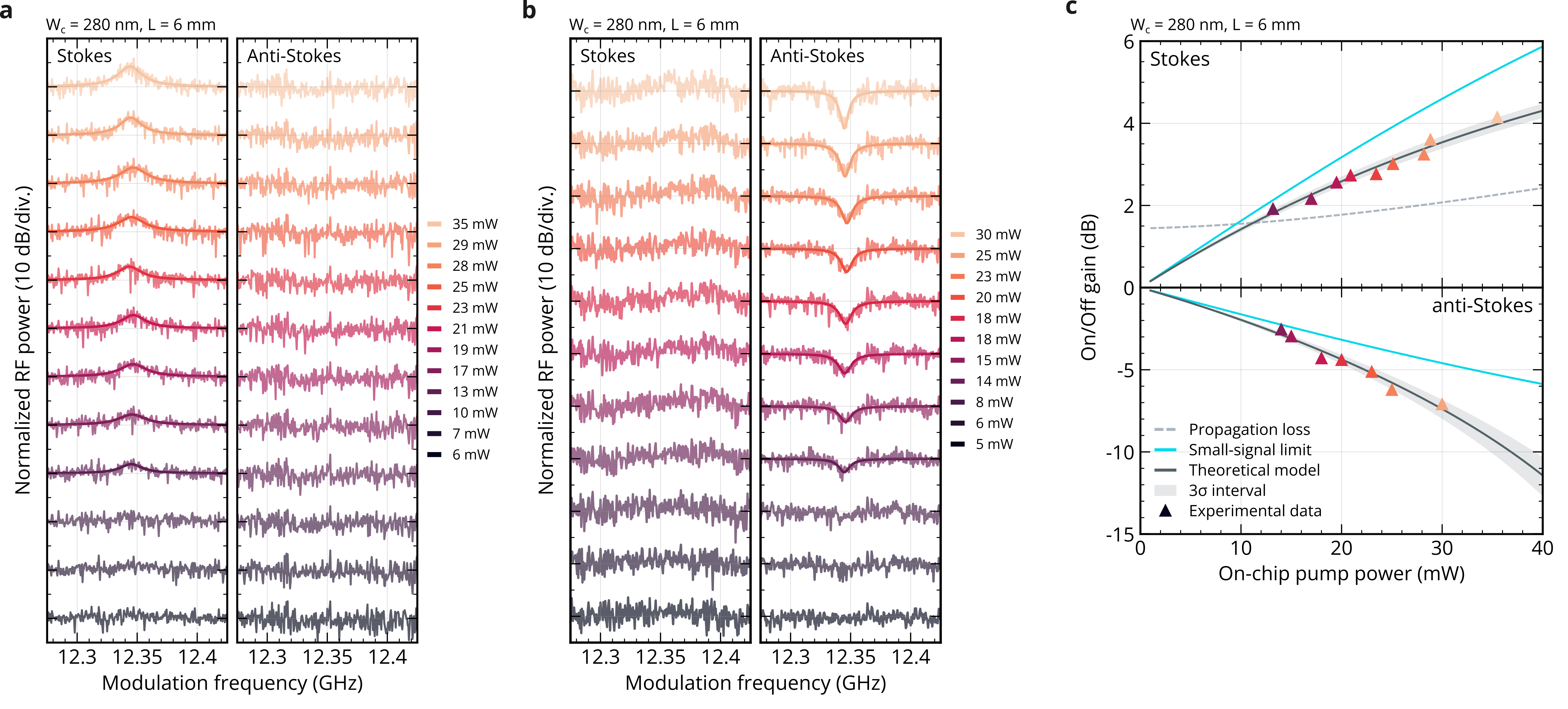}
    \caption{Two-tone pump experiment for the waveguide with core width of $W_\nit{c}=280$ nm and length $L = 6$ mm. \textbf{a}, \textbf{b} Normalized RF beat note between the reference and the Stokes (left panel) and anti-Stokes (right panel) lines as a function of the modulating frequency for different input pump powers when \textbf{a} the anti-Stokes line is filtered out at the input and \textbf{b} the Stokes line is the one filtered out. \textbf{c} On/off gain for the Stokes (upper panel) and anti-Stokes (lower panel) modes at the output of the waveguide as a function of the input pump power. Figures \textbf{c} appear in the main text.}
    \label{fig:2tone_RF}
\end{figure}

Finally, in Fig. \ref{fig:2tone_RF} we show the measured optomechanical response in the two-tone pump experiment we conducted on the 6 mm-long waveguide with core width of $W_\nit{c} = 280$ nm. At low power (the case of our experiment), when only one of the sidebands inputs the waveguides, the generation of the other sideband is not enough to be detected. In order to observe it, a longer waveguide or higher input power would be needed. However, as we show in Fig. \ref{fig:2tone_RF}c and the model predicts (section \ref{sec:3tone_theory}, Fig. \ref{fig:3tone_smallsignal_comparison}), the relative gain for the Stokes line and the relative depletion of the anti-Stokes line do not follow a symmetric behavior. Also, one must note a small red-shift of the mechanical frequency ($\Omega_\m/2\pi = 12.345 \pm 0.001$ GHz, $\Gamma_\m/2\pi = 18 \pm 4$ MHz, $Q_\m \sim690$) compared to the three-tone pump experiment due to a slight degradation of the waveguide after the measurements. 

\begin{table}[H]
    \centering
    \caption{Three-tone pump experiment. Characteristic Brillouin parameters for different geometries. Here, \textit{Sim.} stands for the simulated values. We use the measured $Q_\mathrm{m}$ for the calculations and we assume an optical mode of frequency $\omega_\mathrm{p}/2\pi=193$ THz.}
    \label{tab:3tone_results}
    {\small
    \begin{tabular}{@{}*{8}{c}@{}}
    \toprule
    & & & \multicolumn{2}{c}{$W_\mathrm{c}=280$ nm} && \multicolumn{2}{c}{$W_\mathrm{c}=320$ nm} \\
    \cmidrule{4-5} \cmidrule{7-8}
    & & & 4 mm & 6 mm && 4 mm & 6 mm \\
    \midrule
    \multirow{6}{*}{\begin{sideways} Experiment \end{sideways}} & $\Omega_\mathrm{m}/2\pi$ & [\unit{\GHz}] & $12.358\pm0.003$ & $12.358\pm0.003$ && $11.990\pm0.003$ & $11.988\pm0.004$ \\
    & $\Gamma_\mathrm{m}/2\pi$ & [\unit{\MHz}] & $18\pm1$ & $20\pm1$ && $20\pm1$ & $20\pm1$ \\
    & $Q$ & -- & $\sim690$ & $\sim620$ && $\sim600$ & $\sim600$ \\
    & $G_\mathrm{B}$ (AS) & [\unit{\per\W\per\m}] & $7562\pm130$ & $7211\pm102$ && $6519\pm133$ & $6011\pm 99$\\
    & $G_\mathrm{B}$ (S) & [\unit{\per\W\per\m}] & $7901\pm578$ & $7718\pm453$ && $7137\pm650$ & $6596\pm 335$\\
    & $P_\nit{th}$ & [\unit{\mW}] & 10 & 8 && 11 & 12 \\
    \midrule
    \multirow{2}{*}{\begin{sideways} Sim. \end{sideways}} & $\Omega_\mathrm{m}/2\pi$ & [\unit{\GHz}] & \multicolumn{2}{c}{12.368} && \multicolumn{2}{c}{12.009} \\
    & $G_\mathrm{B}$ & [\unit{\per\W\per\m}] & 7913 & 7110 && \multicolumn{2}{c}{6422}\\
    \bottomrule
    \end{tabular}
    }
\end{table}

\bibliographystyle{unsrt}
\bibliography{references}